\documentclass[11pt]{elsarticle}
\usepackage{lipsum}
\makeatletter
\def\ps@pprintTitle{%
 \let\@oddhead\@empty
 \let\@evenhead\@empty
 \def\@oddfoot{}%
 \let\@evenfoot\@oddfoot}
\makeatother

\usepackage{microtype}
\usepackage{amsmath,amssymb}
\usepackage{setspace}

\usepackage{natbib} 
\setcitestyle{authoryear,open={(},close={)}}
\usepackage{hyperref}
\hypersetup{
colorlinks=true,
linkcolor={blue},
urlcolor={blue},
citecolor={blue}}
\counterwithout{figure}{section}
\usepackage[newfloat,frozencache=true,cachedir=minted-cache]{minted}

\makeatletter
\g@addto@macro\@floatboxreset\centering
\makeatother

\usepackage{xcolor}
\usepackage{graphicx}% Include figure files
\usepackage{bm}% bold math
\usepackage{bbm}% math symbols
\usepackage{braket}

\DeclareUnicodeCharacter{3BC}{$\mu$}
\DeclareUnicodeCharacter{3C3}{$\sigma$}

\usepackage{qcircuit}

\usepackage{listings}

\usepackage{hyperref}% add hypertext capabilities
\newcommand{\custompythonfromfile}[1]{%
\inputminted[
	bgcolor=gray!6,
	fontfamily=tt,
	gobble=0,
	fontsize=\small,
	framerule=0.4pt,
	funcnamehighlighting=true,
	tabsize=2,
	obeytabs=false,
	mathescape=false
	samepage=true, %with this setting you can force the list to appear on the same page
	showspaces=false,
	showtabs =false,
	texcl=false,
	linenos=false,
	xleftmargin=\parindent,
	numbersep=1pt,
	texcomments
]{python}{#1}
}

\newenvironment{custompython}{\VerbatimEnvironment\begin{minted}[
	bgcolor=gray!6,
	fontfamily=tt,
	gobble=0,
	fontsize=\small,
	framerule=0.4pt,
	funcnamehighlighting=true,
	tabsize=2,
	obeytabs=false,
	mathescape=false
	samepage=true, %with this setting you can force the list to appear on the same page
	showspaces=false,
	showtabs =false,
	texcl=false,
	linenos=false,
	xleftmargin=\parindent,
	numbersep=1pt,
	texcomments
]{python}}{\end{minted}}
\graphicspath{{./figs/}}

\begin{document}

%\title{Economics Applications of Monte %Carlo Estimation on a Quantum Computer}
\title{Quantum Monte Carlo for Economics: Stress Testing and Macroeconomic Deep Learning}

\author[a]{Vladimir Skavysh$^{1}$}
\author[a]{Sofia Priazhkina\footnote{Communicating authors. Email addresses: vskavysh@bank-banque-canada.ca, pria@bank-banque-canada.ca
}}
\author[b]{Diego Guala}
\author[b]{Thomas R. Bromley}

\address[a]{Bank of Canada, 234 Wellington Street, Ottawa, ON, K1A 0G9, Canada}
\address[b]{Xanadu, Toronto, ON, M5G 2C8, Canada}

\date{\today}

\begin{abstract}

Computational methods both open the frontiers of economic analysis and serve as a bottleneck in what can be achieved. We are the first to study whether Quantum Monte Carlo (QMC) algorithm can improve the runtime of economic applications and challenges in doing so. We provide a detailed introduction to quantum computing and especially the QMC algorithm. Then, we illustrate how to formulate and encode into quantum circuits (a) a bank stress testing model with credit shocks and fire sales, (b) a neoclassical investment model solved with deep learning, and (c) a realistic macro model solved with deep neural networks. We discuss potential computational gains of QMC versus classical computing systems and  present a few innovations in benchmarking QMC.\\
\end{abstract}


\maketitle

\section{Introduction}

Monte Carlo methods are a group of algorithms that harness randomness to perform a number of computational tasks (\citealp{kalos2008monte}).\footnote{The term was first used in \citet{metropolis1949monte} in the study of differential equations.} This approach is common in various sciences that use statistics, data sampling, and mathematical complexity. When major improvements in computers began to occur, economics was among the first of the social sciences to take advantage of Monte Carlo techniques. However, many numerical economic applications are still not feasible today given classical computational limitations. The primary purpose of this paper is to explore whether the Monte Carlo methodology can be improved with the usage of quantum technology in the context of economic applications and to determine when a quantum advantage can be achieved for these problems. Being pioneers in quantum applications in economics, we also aim to lay down the basic foundation of quantum programming so that economists can benefit from future technological innovations.

The intuition and challenges of Monte Carlo simulations can be grasped by considering a simple problem: calculation of an expectation of a random variable. Simulations of this kind are simple to perform---samples are generated according to the underlying probability distribution, the random variable is then evaluated, and an average is taken. Clearly, the error of the approximation depends on the number of samples generated, resulting in a trade-off between accuracy and time. The trade-off persists even when the sampling is done in parallel on a CPU/GPU cluster due to the computational cost, costs of purchasing hardware and software, and maintenance costs. In practical applications, this trade-off can lead to the Monte Carlo estimation becoming the main
bottleneck in a workflow.

Quantum computations may become a remedy for these problems. Quantum mechanics provides a new class of algorithms that can potentially outperform their classical (non-quantum) counterparts (\citealp{nielsen2010quantum}). These algorithms must be executed on quantum hardware and require development of new technologies to control for excessive noise in computations, referred to as fault-tolerance. 
One algorithm of interest is the so-called quantum Monte Carlo (QMC) algorithm (\citealp{Montanaro2015,xu2018turbulent, Derivatives}), which has the potential to estimate an expectation value with a quadratic speedup (in query complexity) compared to the classical Monte Carlo.\footnote{Note that the algorithm discussed here is conceptually different from the quantum Monte Carlo techniques used to analyze quantum many-body systems (\citealp{pang2016introduction}).} The QMC algorithm adopts established techniques from quantum computing, including amplitude estimation (\citealp{brassard2000quantum}), which we use in this paper.\footnote{Other methods include quantum search, as in \cite{grover1996fast}, and phase
estimation, as in \cite{kitaev1995quantum}.} Implementing QMC requires encoding the problem as a combination of unitary operators (complex matrices that preserve the inner product). These unitaries must be then decomposed into elementary quantum gates---analogs of non-quantum Boolean logical functions---for QMC to be compatible with hardware. Importantly, such decomposition must be efficient for the quadratic speedup to persist. Hence, a major focus has been on identifying efficient encodings of the Monte Carlo problem.\footnote{See \cite{GroverRudolph, Zoufal2019, GroverRudolphProblem} for encoding probability distribution and \cite{Derivatives, vedral1996quantum, herbert2021quantum, woerner2019quantum, Stamatopoulos_2020} for encoding a random variable.}

\begin{sloppypar}
In recent years, the most relevant use-cases developed alongside the QMC algorithm have involved solving
problems in finance, including options
pricing by \cite{Derivatives, Stamatopoulos_2020} and \cite{chakrabarti2021threshold} and risk
analysis by \cite{woerner2019quantum} and \cite{egger2020credit}. However, little work has been carried out in the larger field of economics. More generally, applications of quantum computing to economics have been extremely scarce. Among the few examples of quantum-driven economics known to us are \cite{hull2020quantum} with the general discussion of quantum algorithms in the context of economics and specifically quantum money; \cite{orus2019forecasting} on modeling contagion in financial networks; \cite{alaminos2022quantum} on deep learning techniques for GDP forecasting; \cite{mcmahon2022improving} on improving the efficiency of a large value transfer payments system; \cite{fernandez2022dynamic} on dynamic programming using quantum annealer; and \cite{networks_multiverse} on studying cryptocurrency adoption in networks. This leaves many unexplored opportunities for economists to apply quantum computing in their research and policy. 
\end{sloppypar}

In this work, we focus on two policy-relevant applications in the context of QMC: stress-testing and general equilibrium macro-modeling. For the stress testing, we develop a model where banks experience severe credit losses and engage in fire sales. The credit losses are random and determined by the stress scenario. This setup resembles a typical exercise performed by regulators and central banks to assess financial stability. The goal of the exercise is to evaluate overall capital losses that banks experience following a multi-year stress scenario. We encode this problem onto the quantum circuit and compare the quantum solution with the theoretical prediction. 

For the macro-modeling problem, we solve the stochastic neoclassical investment model using machine (deep) learning. The model is intended to capture relationships between major macroeconomic variables: consumption, capital, production, and productivity shocks. A central part of the deep learning approach is the Monte Carlo estimation of stochastic gradients as in the recent work of \cite{maliar2021deep}. We substitute the QMC algorithm for this part and perform a fair comparison between quantum and classical algorithms by decomposing the problem into elementary gates and calculating the physical runtime. This is the first paper, as far as we are aware, comparing physical runtimes between the two algorithms. We find that QMC can be advantageous over sizable high-performance computing (HPC) clusters. In reaching the computational error of $10^{-8}$, QMC reaches the solution $5.6$ times faster than a strong HPC cluster that an average researcher should be able to rent. If a lower computational error were desired, QMC's advantage would only grow.

We then explore this macroeconomic deep learning application farther by demonstrating QMC's quadratic speedup on a more realistic model. The model contains multiple random variables and is solved with a deep neural network.

In addition to exploring economics problems as a use case for quantum algorithms, one of the key contributions of this work is to provide a detailed investigation into the resources needed to perform QMC. Our investigation requires an efficient decomposition of the algorithm into hardware-compatible quantum gates. This can be done using existing results in the literature that show how to decompose the unitaries that encode the probability distribution~\cite{GroverRudolph, Zoufal2019, GroverRudolphProblem} and random variable of the estimation problem~\cite{Derivatives, vedral1996quantum, herbert2021quantum, woerner2019quantum, Stamatopoulos_2020}. We build upon the state-of-the-art by optimizing a variational quantum circuit with the objective of closely approximating the target probability distribution, using similar methods to those outlined in Ref.~\cite{chakrabarti2021threshold}. Furthermore, we show how a controlled version of the quantum oracle, $\mathcal{Q}$, can be designed without a controlled unitary $\mathcal{F}$, which encodes the classical problem onto the quantum computer.

By combining our variational method for encoding the probability distribution with the decomposition of a linear random variable provided by \cite{egger2020credit} and \cite{Stamatopoulos_2020}, we use the PennyLane software library to simulate the QMC algorithm and monitor its performance and resource requirements as the number of qubits are scaled. The gate depth can be mapped to an algorithmic runtime by assuming each gate is performed within a maximum time, allowing us to compare the runtimes of the classical and quantum Monte Carlo algorithms---an important exercise when gauging the potential for quantum advantage.

We begin in section~\ref{Sec:Econ} by giving an overview of Monte Carlo applications in economics. In section \ref{Notation}, we briefly discuss quantum computation concepts and notation for economists unfamiliar with the subject. We also describe the current state-of-the-art in quantum computing hardware and how soon it might be possible to implement the QMC algorithm for realistic problems on a real quantum device. In section \ref{Sec:QMC}, we provide a mathematical formulation of the quantum Monte Carlo algorithm and describe how it can be performed on a quantum computer. This section is more technical and can be skipped by readers less interested in the details of the algorithm. Section~\ref{Sec:Apps} introduces two economics applications of QMC: (a) stress testing of banks and (b) solving dynamic stochastic general equilibrium (DSGE) models with deep learning. In the latter, we consider two models: a simple one and a more realistic model. In section \ref{Sec:Bench}, we benchmark the QMC solution of the simple DSGE model against the classical Monte Carlo solution. In section \ref{sec:extensions}, we describe the challenges of QMC and how current and future work is likely to address these challenges. We conclude the paper in section \ref{conclusion}. Many mathematical details of the QMC algorithm, how QMC can be decomposed into hardware-compatible gates, and our code implementing the QMC algorithm are all provided in the Appendices.

\section{Monte Carlo Methods in Economics}\label{Sec:Econ}

 Monte Carlo (MC) techniques have broad applicability across the field of economics. We provide a few illustrative examples to define the scope of how QMC could help. 
 
 A common application of MC among economists is to extend the sample size of small datasets. This can be useful when a researcher relies on using methods with statistical properties that only hold asymptotically and when collecting a large sample of the data is expensive or not feasible. A significant part of these kinds of computations involves random sampling. Bootstrapping developed by \cite{efron1979} is a Monte Carlo technique of this sort. A simple version of this method is essentially random draws with replacement. Observations are drawn from a collected dataset with the purpose of creating N alternative data sets. The goal is then to infer the distribution of some statistic of the population data, such as mean or variance. This simple technique can be advanced in many ways (refer, for instance, to \cite{rubin1981bayesian} for the Bayesian analogue). In applied economics, bootstrapping also accompanies regression analysis. For instance, an economist may specify the model as a linear regression and use sampling techniques to infer the properties of the error term and the estimate. The specifics of such procedure and their impact on inference are studied in detail by econometricians (\citealp{hendry1984monte}, and \citealp{davidson1993estimation}, have relevant reviews). 

Economists also use MC for modeling non-standard distributions and calculating their probability measures. As doing so often involves integration, economists are left with numerical methods and simulation techniques as the only remedy. Such techniques can be very sophisticated and go beyond the scope of this paper. While here we focus on independent draws, Markov Chain Monte Carlo methods applied in Bayesian econometrics provide a perfect illustration of how sampling can be done when Monte Carlo draws are auto-correlated (see \citealp{metropolis1949monte}, and \citealp{hastings1970monte}, for the commonly used Metropolis-Hastings algorithm).      

Overcoming the complexity of economic models is another reason for utilizing computational simulations. More and more, economists are relying on realistic modeling assumptions, utilizing big data sets, and including a large number of agents and interactions between them. To solve such models, computational techniques such as MC become indispensable. There are two main sources of complexity in the current models: (i) behavioral or theoretical and (ii) relationship-based or computational. 

The first type of complexity comes from mathematical complexities that economists face when modeling rational behaviors of agents. For instance, the most commonly used class of macroeconomic models, called dynamic stochastic general equilibrium models (DSGE), cannot be solved explicitly apart from some trivial cases, and require fixed-point convergence algorithms for equilibrium search. Because the behaviors of agents in such models are assumed to be interdependent across time and economic sectors, these models become computationally challenging even with little heterogeneity of actions and agents. Thus, computational methods, such as MC techniques, are essential and receive much consideration from macroeconomists (e.g., see \cite{judd1998numerical} for the chapters on MC methods). Later in this paper, we show how machine-learning techniques can be applied to solve a DSGE model and where QMC could play a pivotal role.

The second source of complexity arises in the agent-based models (ABM) and other computational economics models (\citealp{hommes2006heterogeneous} provides a comprehensive review). Unlike in DSGE models, agents now behave myopically according to simple heuristics. Thus, the researchers are less focused on finding the equilibrium fixed point. However, the number of agents is often large, and the computational demands grow rapidly as the network of interactions between agents grows in size. Models of this kind are helpful for including bounded-rationality and heterogeneity among agents. Thus, they are less predictable in terms of computational output and intrinsically rely on simulations. This creates even more demands for hardware performance, as a researcher may need to run the model many times to test various calibrations and model treatments. We use an approach similar to ABM for our second example when discussing stress testing of banks.   

Finally, Monte Carlo methods are essential for reducing the mathematical complexity of many economic models. The instances of this kind are too many to list. Economists tend to rely on numerical methods for calculating multidimensional integrals, finding fixed points, solving differential equations, etc. In finance alone, the applications range from option pricing to optimal portfolio selection (see for instance \citealp{boyle1997monte}). As such, overcoming the computational costs would improve the modeling abilities of economists.

\section{Quick Introduction to Quantum Computing for Economists}\label{Notation}
\subsection{Quantum advantage}
Quantum computing applies the laws of quantum mechanics to perform computations. Quantum mechanics is a theory describing microscopic systems with low gravity and speeds much lower than the speed of light.  Within quantum computation, numerous algorithms have been developed with proven theoretical speedup over the best available classical algorithms (\citealp{nielsen2010quantum,Zoo}). As quantum hardware improves, this speedup should be realised in practice. This will usher in the age of “quantum advantage" (or “supremacy") where even the best classical supercomputers cannot compete against quantum computers for certain problems. Although such advantage for useful problems is yet to be realized, quantum advantage has been claimed, for a few esoteric problems (\citealp{arute2019quantum,zhong2020quantum,madsen2022quantum}). 

However, the initial quantum advantage in practice might actually come not in terms of computational speed, but rather in terms of financial cost or energy efficiency (and, therefore, carbon emissions). Current virtual machine prices suggest that quantum computing could become cost effective beyond around $1000$ to $10000$ CPUs.\footnote{See, for example,
\href{https://learn.microsoft.com/en-us/azure/quantum/pricing}{https://learn.microsoft.com/en-us/azure/quantum/pricing} and \href{https://azureprice.net}{https://azureprice.net}.} Given current power consumption of quantum computers, such as Google's Sycamore which uses $26kW$ of power (\citealp{arute2019quantum}), quantum computing may be more energy efficient than roughly $1000$ CPUs. Thus, even in cases where quantum speedup is difficult to realize relative to a classical supercomputer, quantum computing could still prove advantageous in terms of financial cost and energy efficiency. In this paper, we only focus on computational scalability of QMC vs MC. Nevertheless, readers interested in cost or energy usage could easily extend our results to these cases.

\subsection{Qubit, superposition, measurement, and entanglement}

The basic unit of quantum computation is a \textit{qubit}. Qubit is a quantum counterpart of a bit in classical computation. A classical bit, realized with the transistor, can have the value of either 0 or 1. By contrast, a quantum bit, which can be realized with a superconductive loop, trapped ion, photonics, or other methods, is a linear combination or \textit{superposition} of these two states. Mathematically, a qubit can be described as $a\ket{0}+b\ket{1}$, where $a$ and $b$ are complex constants and $\ket{0}$ and $\ket{1}$ refer to states with values $0$ and $1$, respectively. 

\begin{figure}[h]
\includegraphics[width=4cm]{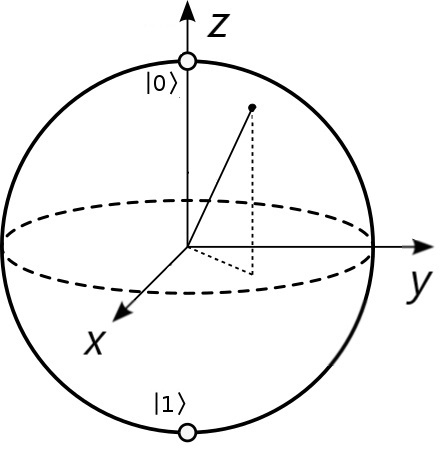}
\caption{Geometric representation of a qubit as a complex vector of length one. Poles of the sphere correspond to observable states $|0\rangle$ and $|1\rangle$.}
\label{bloch}
\end{figure}

Superposition can be understood geometrically. A geometrical representation of a qubit is a radius vector inside a unit sphere (see Figure \ref{bloch}). Poles of the sphere correspond to the basis states of the qubit: $\ket{0}$ and $\ket{1}$. A qubit is in a superposition whenever it is not in the basis states.

Whereas a qubit may initially exist in a superposition of states $\ket{0}$ and $\ket{1}$, the process of \textit{measuring} the qubit forces the qubit to assume the value of either $\ket{0}$ or $\ket{1}$. Qubits tend to be measured at the end of a quantum algorithm to obtain the output. To explain the intuition behind the superposition and measurement, consider the famous thought experiment of Erwin Schrödinger: place a cat with a deadly radioactive element in a sealed box. In the quantum world, the cat is both dead and alive before the box is opened. It is only when the state of the cat is measured (when the box is opened) that the cat becomes “dead" or “alive."

Another important concept of quantum mechanics is \textit{entanglement} described by Albert Einstein as a “spooky action at a distance." When two qubits are linked together, or \textit{entangled}, action applied to one of the qubits immediately impacts the other qubit. If two qubits are entangled such that their spins are reversed, measurement of $\ket{0}$ for the first qubit would correspond to the measurement of $\ket{1}$ for the second one.   

Superposition and entanglement make it possible to manipulate an exponentially large number of states simultaneously. Whereas a single qubit is a superposition of $2=2^1$ states, using $N$ qubits enables one to manipulate $2^N$ states simultaneously. As a result, with only 300 qubits, it is possible to operate on more states than there are number of atoms in the universe. This exponential increase in the computational power with each additional qubit is what brings computational advantage to quantum computers over classical machines. 

\subsection{Bra-ket notation}
In quantum physics, quantum states are defined to be complex unit vectors in a Hilbert space---a vector space with an inner product that defines the distance between two vectors.

For the convenience of linear algebra notations, physicists use angle brackets (“bra-ket" notation). In particular, these brackets can be used to distinguish a vector space from its dual vector space. For instance, an arbitrary qubit can be presented as a linear combination of the two basis states,
\begin{equation}
\ket{\psi} = \alpha \ket{0} + \beta \ket{1} ,
\label{qub}
\end{equation}
where  $\alpha$ and $\beta$ are complex numbers.
For vector $\ket{\psi}$, there is a corresponding covector in the dual space 
\begin{equation}
\bra{\psi} = \alpha^* \bra{0} + \beta^* \bra{1},
\end{equation}
where $\alpha^*$ and $\beta^*$ are conjugate scalars to $\alpha$ and $\beta.$
 To preserve probability, quantum states are taken to be unit vectors. Therefore, the inner product of a vector and its covector yields 
\begin{equation}
\bra{\psi}\psi \rangle = {\alpha}^* \alpha + {\beta}^* \beta =1.
\end{equation}

It might be more familiar to think of the bra $\bra{\psi}$ as a row vector and $\ket{\psi}$ as a column vector. Thus, the product $\bra{\psi}\phi \rangle$ of any two quantum states $\ket{\phi}$ and $\bra{\psi}$ returns a scalar while the product $\ket{\phi}\bra{\psi}$ returns a matrix. 

Any state of a qubit can be represented as a linear combination of the basis states. In the standard basis, the qubit will assume a value of either $\ket{0}$ or $\ket{1}$ when measured. However, an infinite number of other bases can be defined and used for measuring qubits. A notable example is the Hadamard or x-basis: $$\ket{+}=\frac{\ket{0}+\ket{1}}{\sqrt{2}}$$ $$\ket{-}=\frac{\ket{0}-\ket{1}}{\sqrt{2}}.$$

We can apply usual linear algebra operations to quantum states. Moreover, the definitions above can be extended to systems of multiple qubits. We will be using the outer product $\otimes$ whenever we speak about the joint state of multiple qubits. For simplicity, we will write:
$$
\ket{0}^{\otimes m} = \ket{0}\otimes ...\otimes \ket{0}
$$
for $m$-dimensional qubit system. Note that an $m$-dimensional qubit system has $M=2^m$ basis vectors, which we will denote as: 
$$
\ket{1},\ket{2},...,\ket{M}.
$$

\subsection{Quantum gates and unitary operators}

\textit{Quantum gate} is an operation that changes a state of a qubit. It is analogous to a non-quantum Boolean logical function. Current quantum computers can implement around a dozen different quantum gates. Whenever a problem is executed on a quantum computer, it needs to be represented as a mathematical composition of the quantum gates. We define several quantum gates that we use to implement the QMC algorithm.

\textit{RX, RY, RZ gates}---gates that rotate the state of a qubit around axes X, Y, Z by a given angle (refer to the sphere representation of a qubit).

\textit{CNOT (“Controlled NOT") gate}---a two-qubit gate that flips the second qubit from $ |1\rangle$ to $|0\rangle$ or from $ |0\rangle$ to $ |1\rangle$  if and only if the first qubit is $ |1\rangle$. For instance, 
$$
CNOT( |1\rangle \otimes|1\rangle)=|1\rangle\otimes|0\rangle
$$
$$
CNOT( |0\rangle \otimes|1\rangle)=|0\rangle\otimes|1\rangle.
$$

\textit{Hadamard gate}---a gate that transforms either computational basis state into equal superposition of the basis states. That is, after the transformation, the probability of measuring $|0\rangle$ or $|1\rangle$ become equal to 50\% each:
$$
 \textstyle |0\rangle \mapsto \frac {|0\rangle +|1\rangle }{\sqrt {2}}\textstyle
 $$
 $$
\textstyle |1\rangle \mapsto \frac {|0\rangle -|1\rangle }{\sqrt {2}}\textstyle. 
$$

Later in the paper, we will often refer to a \textit{quantum circuit}---a sequence of quantum gates, measurements of output qubits, and other actions needed for a given task to be executed. In translating a theoretical problem onto real quantum hardware, the challenge is minimizing 
\textit{circuit depth}, the number of sequential gate executions necessary to run the quantum circuit. Within the quantum circuit, a collection of qubits assigned to some computational task are referred to as a \textit{register} (of qubits). Meanwhile, \textit{ancilla qubits} serve more of a supporting role, enabling some specific computational goal within the circuit, such as storing the expectation value of some random variable.

In quantum computation, mathematical tasks are often performed with the help of \textit{unitary operators}, complex matrices that preserve the inner product. The reason why unitary operators are applied to quantum states is that quantum states (unit vectors) remain unit vectors (valid quantum states) under unitary transformations. In practice, hardware is set up such that the each qubit is initialized in the zero state, $\ket{0}$. As such, it is sufficient to only describe how a unitary operates on the $\ket{0}$ state whenever a new qubit is introduced in the computation.

\subsection{Transition between quantum states}

Mathematical representation of the quantum world can be linked to experimental data. For a qubit in an arbitrary state $\ket{\psi}=\alpha\ket{0}+\beta\ket{1}$, the probability of measuring the qubit to be in state $\ket{0}$ is $|\bra{0}\psi \rangle|^2=\alpha^2$. More generally, the probability of transitioning from state $\ket{\psi}$ to state $\ket{\phi}$ is $|\bra{\phi}\psi \rangle|^2$.

For gaining more intuition about unitaries and probabilities, consider unitary
$\mathcal{A}$ defined as the operator that performs transformation of initial state $\ket{0}$ to state
\begin{equation}\label{A_unitary}
\mathcal{A}\ket{0} = \sqrt{p(0)}|0\rangle +  \sqrt{p(1)}|1\rangle.
\end{equation}

This unitary encodes the probability distribution $p(\cdot)$ defined on the basis states. The probability of transitioning from the output state 
$\mathcal{A}\ket{0}$ to
the basis state $\ket{0}$ is exactly $p(0)$, and the probability of arriving from the output state 
$\mathcal{A}\ket{0}$ to
 the basis state $\ket{1}$ is exactly $p(1)$.

\subsection{Current quantum computing hardware} \label{Sec:q_hardware}

\begin{figure}[t]
\includegraphics[width=0.85\textwidth]{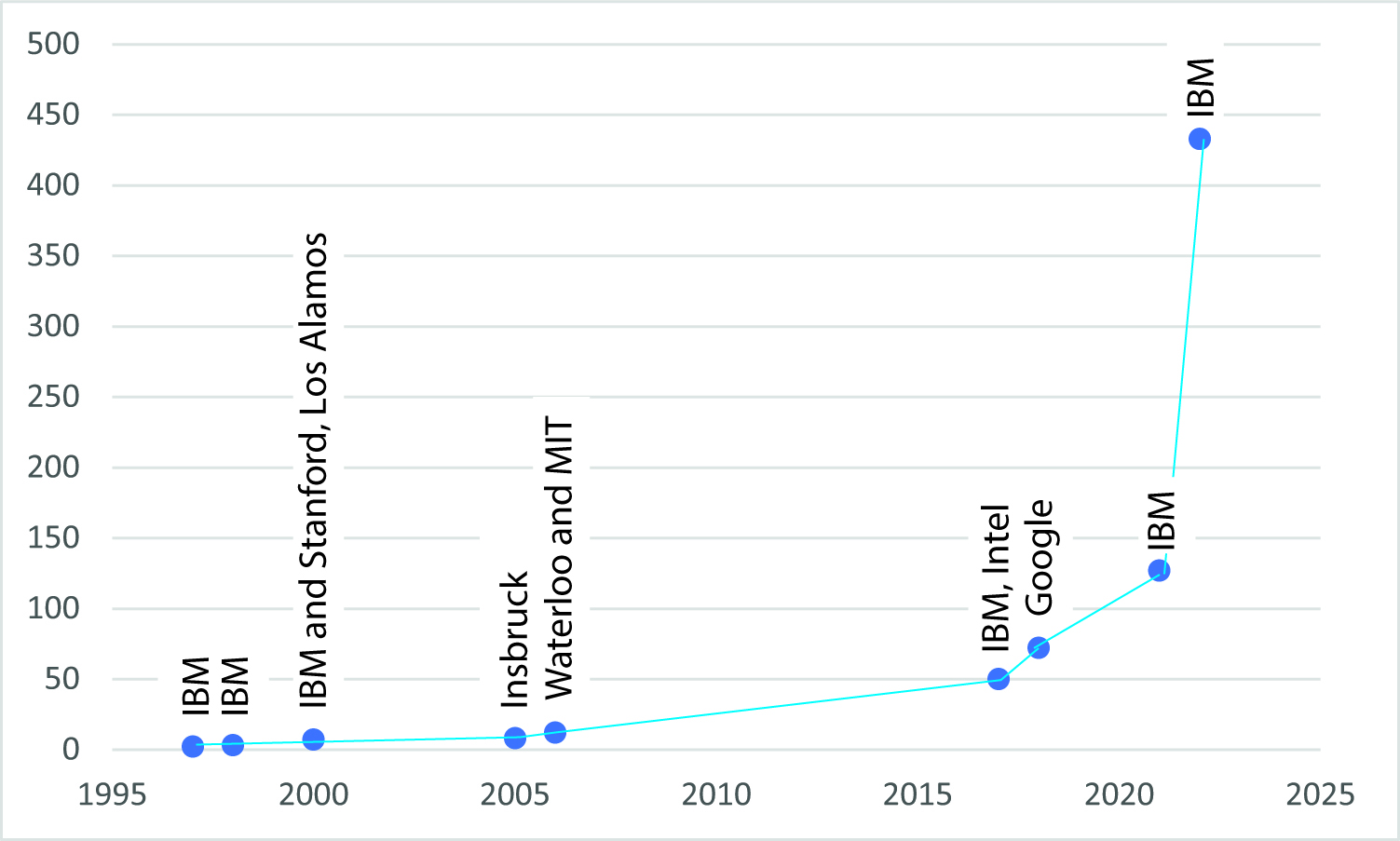}
\caption{Progress in quantum hardware over time. Quantum computers are growing exponentially in the number of qubits, currently doubling approximately once every one or two years.}
\label{fig:qubits_time}
\end{figure}

Progress in quantum hardware can be visualized by plotting how the number of qubits (as part of a single quantum machine) has grown over time. As can be seen in Figure \ref{fig:qubits_time}, the number of qubits is currently doubling roughly once every year or two.

In terms of the qubits themselves, they can be realized using many different physical systems, each having its own advantages and disadvantages. At present, the leading systems are based on superconducting qubits (\citealp{kjaergaard2020superconducting}) and trapped ion qubits (\citealp{bruzewicz2019trapped}). However, systems with alternative technologies, such as photonics (\citealp{flamini2018photonic}), semiconductor qubits, (\citealp{chatterjee2021semiconductor}), neutral atom qubits (\citealp{henriet2020quantum}), and topological qubits (\citealp{oreg2020majorana}), are also undergoing rapid development and could well overtake the current leaders.\footnote{Additional types of quantum hardware systems exist, most notably quantum annealers. However, these systems (at least in their current form) cannot execute the QMC algorithm, and so are not reviewed here. Interested reader may consult \cite{yarkoni2022quantum} for a review of quantum annealing and \cite{mcmahon2022improving} for an example of how quantum annealing might be used to improve efficiency of a payments system.}

To characterize the performance of qubits, a few key metrics are usually sufficient. Coherence time, also known as the $T_1$ time, describes how long a qubit can retain information (higher is better). Gate time (or gate duration) measures how long it takes to implement a given (one-qubit or two-qubit) quantum gate (shorter is better). Finally, fidelity measures (as a percentage) how well a given quantum gate achieves its intended function (higher is better).

Superconducting qubits have had a head start because they could leverage preexisting technologies from classical electronics. These qubits have fast gate times of $\mathcal{O}(10^{-8}s)$ to $\mathcal{O}(10^{-7}s)$ (\citealp{kjaergaard2020superconducting}) and high fidelity of $99.8\%$ and $99.5\%$ for single- and two-qubit gates, respectively (\citealp{jurcevic2021demonstration}). However, they have short coherence times of $\mathcal{O}(10^{-5}s)$ to $\mathcal{O}(10^{-4}s)$ and require extreme cryogenics cooling. Also, problems such as surface noise and qubit-to-qubit variation present challenges to scalability of superconductive qubit systems (\citealp{leon2021materials}). Nevertheless, superconductive qubit hardware has been progressing rapidly in recent years, with the largest computer currently having $433$ qubits.\footnote{\href{https://newsroom.ibm.com/2022-11-09-IBM-Unveils-400-Qubit-Plus-Quantum-Processor-and-Next-Generation-IBM-Quantum-System-Two}{https://newsroom.ibm.com/2022-11-09-IBM-Unveils-400-Qubit-Plus-Quantum-Processor-and-Next-Generation-IBM-Quantum-System-Two}}

Trapped ion qubits have even higher gate fidelities ($99.9999\%$ and $99.9\%$ for single- and two-qubit gates, respectively, \citealp{harty2014high,ballance2016high}) and unmatched coherence times (unlimited for any practical purpose, \citealp{leon2021materials}). On the other hand, their gate times are slower (on the order of $\mathcal{O}(10^{-6}s)$ to $\mathcal{O}(10^{-4}s)$, \citealp{ballance2016high}), they require ultra-high vacuum, and it is difficult to scale them due to challenges in controlling the qubits (\citealp{leon2021materials}). Currently, the largest trapped ion computer has $32$ qubits.\footnote{\href{https://ionq.com/posts/may-17-2022-ionq-forte}{https://ionq.com/posts/may-17-2022-ionq-forte}}

In section \ref{Sec:Bench}, we demonstrate that fast gate times are essential for QMC to achieve quantum advantage in practice. Thus, it is important to note that gate times less than $1$ns have already been achieved (\citealp{he2019two}) and that further speed improvements are likely in the future.

\subsection{Fault-tolerance}\label{sec:fault_tol}

Even with gate fidelities as high as $99.9\%$, the result of a computation will be reduced to noise after only several hundred gates. This is not sufficient to execute algorithms such as QMC (as is evident from results in section \ref{Sec:Bench}). Thankfully, there exists the quantum fault-tolerance theorem (or threshold theorem) which shows that computational errors can be ``corrected" to arbitrarily low levels, given enough qubits and that the qubits have fidelities beyond some threshold (\citealp{shor1996fault,aharonov1997fault,knill1998resilient,kitaev2003fault}). In practice, it is believed that fidelities above $99\%$ should be sufficient (\citealp{fowler2009high}). As for the number of qubits, only millions of qubits are likely to be ``enough." For example, \cite{gidney2021factor} shows that to factor $2048$ bit RSA integers within $8$ hours, some $20$ million qubits are necessary with fidelities of $99.9\%$. Assuming that the rate of qubit growth remains as in the past few years (and that the fidelities remain beyond the necessary threshold), a million qubits should be attainable within $10$ to $15$ years. Either way, error-correction has already been demonstrated experimentally (\citealp{acharya2022suppressing}). This suggests that building useful quantum computers might now be mainly a technical challenge, and is, therefore, a matter of time.

\subsection{Access to quantum computers}

Many quantum computers, including many state-of-the-art systems, are available for anyone to use. For entry-level experimentation, IBM-Q offers free access to machines with seven or fewer qubits.\footnote{\href{https://quantum-computing.ibm.com/}{https://quantum-computing.ibm.com/}} Many other machines can be rented via the cloud (through services such as Amazon Bracket, Azure Quantum, or Xanadu Cloud) or by partnering with various quantum companies, startups, or academics. This includes Xanadu's Borealis, which achieved quantum advantage in 2022 (\citealp{madsen2022quantum}) and will soon include IBM's 433-qubit Osprey computer.\footnote{See \href{https://www.xanadu.ai/products/borealis/}{https://www.xanadu.ai/products/borealis/} and \href{https://quantum-computing.ibm.com/services/resources/docs/resources/manage/systems/processors\#osprey}{https://quantum-computing.ibm.com/services/resources/docs/resources/manage/systems/processors\#osprey}.}

With the notable exception of Google, companies developing quantum hardware have mostly followed the strategy of making their quantum computers available to the public. This is likely to continue in the near future to encourage early use case exploration and to grow the user base capable of working with the computers. However, once quantum advantage is realized for useful problems, the law of supply and demand might make other business strategies preferable for the hardware companies. In the meantime, accessing quantum machines is a matter of cost or academic and business connections. 

\subsection{PennyLane library}

Throughout this paper,
code blocks are provided to show how relevant parts of the algorithm can be implemented using
the PennyLane software library (\citealp{bergholm2018pennylane}). PennyLane is an open-source library with thousands of contributors from around the world. The implementation of the QMC algorithm within PennyLane has been our contribution to the library. PennyLane's documentation and installation instructions are available at \href{https://pennylane.ai}{pennylane.ai}.

\section{Quantum Algorithm for Monte Carlo Simulations of Moments}\label{Sec:QMC}

This section provides a review of Monte Carlo (MC) estimation and how it can be
performed on a quantum computer. Starting from the basics of MC estimation, we show
how the problem can be encoded as a quantum algorithm and then subsequently sped up using
amplitude estimation (\citealp{brassard2000quantum,Montanaro2015,xu2018turbulent, Derivatives}).

\subsection{Classical Monte Carlo sampling}

In practice, Monte Carlo draws are often used to evaluate moments of a random variable. We show how this can be done on a quantum computer using mean function as an example.  Consider a random variable $f(\bm{x})$, defined as a measurable function of
$\bm{x}$ specified on $\mathbb{R}^{d}$ with a probability density function $p(\bm{x})$. 
The expectation value of $f(\bm{x})$ can be written as
\begin{equation}\label{Eq:expectation_val}
\mu := {\rm E}[f(\bm{x})] = \int_{\mathbb{R}^{d}} p(\bm{x})f(\bm{x}) \,\,\, d\bm{x}.
\end{equation}
One approach to approximating ${\rm E}[f(\bm{x})]$ is using MC estimation. Here and later in the paper, we assume that $f(x)$ is well-behaved with finite values and non-zero variance. To approximate ${\rm E}[f(\bm{x})]$, we randomly sample $N$ points $\bm{x}_{i}$ according to density $p$ and calculate the average:
\begin{equation}
\hat{\mu} = \frac{1}{N} \sum_{i=1}^{N} f(\bm{x}_{i}).
\end{equation}

The estimate has an absolute error $|\mu - \hat{\mu}|$. The probability that this error is larger than a fixed $\varepsilon > 0$ can be upper bounded using Chebyshev's inequality as
\begin{equation}
\mbox{Pr}\left(|\mu - \hat{\mu}| \geq \varepsilon\right) \leq \frac{\sigma^{2}}{N \varepsilon^{2}},
\end{equation}
where $\sigma$ is the standard deviation of $f(\bm{x})$ (see \citealp{Montanaro2015}, for details). Hence, for a constant probability, we set
\begin{equation}\label{sqrt_root_classical}
N \propto \frac{1}{\varepsilon^{2}}.
\end{equation}
Therefore, the number of samples we must generate is inversely proportional to the square of the target error. Next, we will see how the number of unitaries required in QMC scales as $1 / \varepsilon$, that is, a quadratic speedup relative to that above.

\subsection{Quantum Monte Carlo sampling}

We now show how the Monte Carlo estimation algorithm can be carried out using a qubit-based
quantum circuit. Due to the binary nature of a qubit, we first discretize the %continuous-valued
problem into a space $X$ consisting of $M$ grid points, with probability
mass function $p(i)$ and random variable $f: X \rightarrow [0, 1]$, so that the objective is to
measure the expectation value
\begin{equation}
\mu = \sum_{i\in X} p(i) f(i).
\end{equation}
As before, Monte Carlo can be used to provide an estimate $\hat{\mu}$. Discretization of
continuous-valued problems is an established area with a variety of
approaches to define $p(i)$ above (\citealp{Chakraborty2015, chakrabarti2021threshold, Derivatives}). Note that discretization may introduce an error $\varepsilon_{\rm disc}$ that should be kept below the Monte
Carlo error $\varepsilon$.

We restrict our problem to a random variable that maps to the
interval $[0, 1]$ so that the problem is compatible with encoding into a quantum circuit. When the maximum $f_{max}$ and minimum $f_{min}$ of the function are available for the discrete space $X$, this
restriction can be achieved by renormalizing $f(\bm{x})$ as
\begin{equation}
    f_{norm}(\bm{x}) = \frac{f(\bm{x})-f_{min}}{f_{max}-f_{min}}.
\end{equation}

For building a quantum alternative of the problem, we define two unitary operators: unitary $\mathcal{A}$ for modeling function $p(\cdot)$ and unitary $\mathcal{R}$ for modeling function $f(\cdot)$ in equation (\ref{Eq:expectation_val}).   

For defining $\mathcal{A}$, consider a register of $m$ qubits, so that the grid size we selected is $M = 2^{m}$. We define unitary
$\mathcal{A}$ as the one that performs the transformation of an initial state $\ket{0}^{\otimes m}$ to a state
\begin{equation}\label{A_unitary2}
\mathcal{A}\ket{0}^{\otimes m} = \sum_{i\in X} \sqrt{p(i)}|i\rangle.
\end{equation}
This unitary encodes the probability distribution $p(\cdot)$ because the probability of arriving from the output state 
$\mathcal{A}\ket{0}^{\otimes m}$ to
 basis state $\ket{i}$ is exactly $p(i)$.
 
 The random variable $f(\cdot)$ is
encoded by adding
an additional \textit{ancilla} qubit to output $i=\mathcal{A}\ket{0}^{\otimes m}$ and applying
another unitary $\mathcal{R}$ that performs the transformation
\begin{equation}
\mathcal{R} |i\rangle |0\rangle= |i\rangle \left(\sqrt{1 - f(i)} |0\rangle + \sqrt{f(i)}|1\rangle\right).
\end{equation}

Similarly, this unitary encodes the probability distribution $f(\cdot)$, because the probability of measuring the ancilla qubit in the state $\ket{1}$ is equal to $f(i)$.

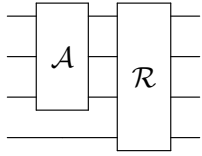
\begin{figure}[t]
\mbox{
\Qcircuit @C=1em @R=0.5em {
& \multigate{2}{\mathcal{A}} & \multigate{3}{\mathcal{R}} & \qw \\
& \ghost{\mathcal{A}} & \ghost{\mathcal{R}} & \qw \\
& \ghost{\mathcal{A}} & \ghost{\mathcal{R}} & \qw \\
& \qw & \ghost{\mathcal{R}} & \qw \\
}
}
\caption{The $\mathcal{F}$ unitary for performing a Monte Carlo estimation. Here, the
$\mathcal{A}$ unitary encodes a $2^{m}$-dimensional probability distribution using $m=3$ qubits
and the $\mathcal{R}$ unitary encodes the expectation value $\mu$ onto the ancilla qubit.}
\label{fig:F_unitary}
\end{figure}

The expectation value $\mu$ can then be encoded onto the ancilla qubit by combining the
$\mathcal{A}$ and $\mathcal{R}$ unitaries together according to
\begin{equation}
\mathcal{F} := \mathcal{R} \left(\mathcal{A} \otimes \mathbbm{1}_{2}\right),
\end{equation}
as shown in Figure~\ref{fig:F_unitary}.

Then the output state of the Monte Carlo problem is defined as
\begin{align}
\label{chi}
\ket{\chi} &:= \mathcal{F} \ket{0}^{\otimes m + 1} \notag \\ &= \sum_{i \in X} \sqrt{p(i)} \,\,\,|i\rangle \otimes \left(\sqrt{1 - f(i)} |0\rangle + \sqrt{f(i)}|1\rangle\right).
\end{align}
The probability of measuring the ancilla qubit in state $\ket{1}$ is
given by
\begin{equation}
P_{1} = \langle \chi | \left(\mathbbm{1}^{\otimes m} \otimes |1\rangle\langle1|\right) | \chi \rangle = \sum_{i \in X}p(i) f(i) = \mu.
\end{equation}

In practice, we cannot obtain the probability $P_{1}$ exactly and hence the expectation value
$\mu$. Instead, we need to sample from the circuit multiple times, resulting in a binary string from which
the probability of observing $|1\rangle$ can be inferred.
Suppose we perform $N$ measurements of the
ancilla qubit and calculate an estimate $\hat{P}_{1} = \hat{\mu} = N_{1} / N$, with $N_{1}$ being defined as the
number of times $|1\rangle$ was measured. Since we are performing Bernoulli trials, the variance of our
estimate will be
\begin{equation}
{\rm Var} (P_{1}) =: \varepsilon^{2} = \frac{P_{1} (1 - P_{1})}{N}.
\end{equation}
Hence, the number of trials scales with an inverse-squared relationship relative to the standard deviation $\varepsilon$:
\begin{equation}
N \propto \frac{1}{\varepsilon^{2}}.
\end{equation}
This approach has successfully encoded the Monte Carlo estimation problem into a quantum
circuit, but provides no speedup.

\subsection{Applying amplitude estimation} \label{ApplyingAmplitude}
In the previous section, we have encoded the Monte Carlo problem onto a register of qubits. In this section, we show how a speedup can be obtained by applying the amplitude estimation algorithm (\citealp{brassard2000quantum}) and using an
additional register of qubits to store the result.

Consider the unitary operator
$\mathcal{V} = \mathbbm{1}_{m+1} - 2 \mathbbm{1}_{m} \otimes |1\rangle\langle 1|$ applied to the output of the Monte Carlo circuit we have so far.\footnote{The important property of this operator is that it is 
Hermitian, meaning that the transpose and  complex conjugate of the operator will return the operator itself.} The unitary nature of $\mathcal{V}$ allows us to write
\begin{equation}
\mathcal{V}| \chi \rangle = \cos\left(\pi \theta\right)|\chi\rangle + e^{i \phi} \sin\left(\pi \theta\right)|\chi^{\perp}\rangle,
\end{equation}
with $\theta$ and $\phi$ being angles and $|\chi^{\perp}\rangle$ being a state orthogonal to $\ket{\chi}$.
Hence, measuring the expectation value of $\mathcal{V}$ for the state
$\ket{\chi}$ gives
\begin{equation}\label{Eq:Theta}
\langle \chi | \mathcal{V} | \chi \rangle = \cos \left(\pi \theta\right) = 1 - 2 \mu.
\end{equation}
If the angle $\theta$, referred to as the ``phase," was known to us, we would use the last result to solve the Monte Carlo problem, meaning we would derive the expected value
\begin{equation}
\label{mu_theta}
\mu = \frac{1-\cos(\pi\theta)}{2}.   
\end{equation}

The phase $\theta$ can be found by specifying another unitary $\mathcal{Q}$ with eigenvalues $e^{\pm 2 \pi i \theta}$. This can be achieved with the help of Grover's diffusion
operators (\citealp{grover1996fast}) that perform a rotation in the subspace spanned by
$\ket{\chi}$ and $\ket{\chi^{\perp}}$. Following \cite{brassard2000quantum} and \cite{ Derivatives},
it can be shown that
\begin{equation}\label{Eq:Q}
\mathcal{Q} = (\mathcal{F}\mathcal{Z}\mathcal{F}^{\dagger} \mathcal{V})^{2},
\end{equation}
where $\mathcal{Z} = \mathbbm{1}_{m + 1} - 2 (\ket{0}\bra{0})^{\otimes m + 1}$.\footnote{Moreover, it
holds that $\ket{\chi} = (\ket{\chi^{+}} + \ket{\chi^{-}})/ \sqrt{2}$, where $\ket{\chi^{\pm}}$
are the eigenstates of $\mathcal{Q}$ with eigenvalues $e^{\pm 2 \pi i \theta}$,
respectively (\citealp{xu2018turbulent}).}

\begin{figure}[t]
\mbox{
\Qcircuit @C=1em @R=0.5em {
& \multigate{3}{\mathcal{F}} & \multigate{3}{\mathcal{Q}^{2^{n - 1}}} & \multigate{3}{\mathcal{Q}^{2^{n - 2}}} & \qw & \cdots & & \multigate{3}{\mathcal{Q}} & \qw & \qw  \\
& \ghost{\mathcal{F}} & \ghost{\mathcal{Q}^{2^{n - 1}}} & \ghost{\mathcal{Q}^{2^{n - 2}}} & \qw & \cdots & & \ghost{\mathcal{Q}} & \qw & \qw  \\
& \ghost{\mathcal{F}} & \ghost{\mathcal{Q}^{2^{n - 1}}} & \ghost{\mathcal{Q}^{2^{n - 2}}} & \qw & \cdots & & \ghost{\mathcal{Q}} & \qw & \qw  \\
& \ghost{\mathcal{F}} & \ghost{\mathcal{Q}^{2^{n - 1}}} & \ghost{\mathcal{Q}^{2^{n - 2}}} & \qw & \cdots & & \ghost{\mathcal{Q}} & \qw & \qw \\
& & & & & & & & &\\
& & & & & & & & &\\
& & & & & & & & &\\
& \gate{H} & \ctrl{-4} & \qw & \qw & \cdots & & \qw & \multigate{4}{\rm{QFT}^{-1}} & \qw \\
& \gate{H} & \qw & \ctrl{-5} & \qw & \cdots & & \qw & \ghost{\rm{QFT}^{-1}} & \qw\\
& \vdots & & & & \ddots & & & \\
& & & & & & & &\\
& \gate{H} & \qw & \qw & \qw & \cdots & & \ctrl{-8} & \ghost{\rm{QFT}^{-1}} & \qw
}
}
\caption{The QMC algorithm applies phase estimation for the unitary $\mathcal{Q}$ onto an input
state prepared by $\mathcal{F}$ using $n$ phase estimation qubits (bottom half of circuit). By
sampling the phase estimation qubits in the standard basis, the eigenvalues of $e^{\pm 2 \pi i \theta}$ of $\mathcal{Q}$ can be estimated.}
\label{fig:QMC}
\end{figure}
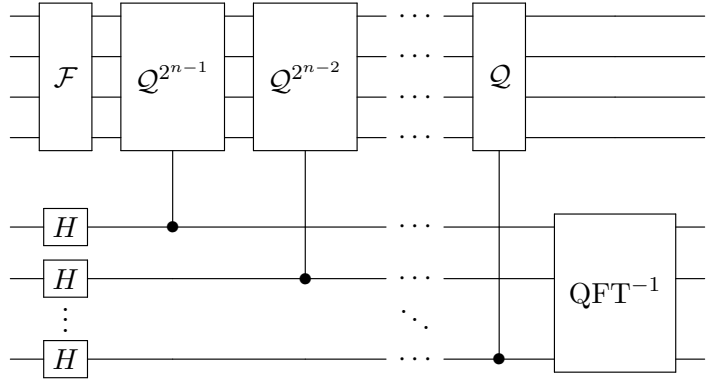

With the help of the quantum phase estimation
algorithm (\citealp{kitaev1995quantum}), we can estimate $\theta$ as shown in Figure~\ref{fig:QMC}. The algorithm introduces an additional register of $n$ qubits (“phase estimation qubits"). These qubits control the application of the $\mathcal{Q}$ operator, which has been raised to various powers of $2^k$ (where $k=1\dots n-1$).\footnote{This process is known as “phase kickback" (\citealp{cleve1998quantum}).} The estimation qubits are in states of a Fourier basis. To return to the standard basis composed of $\ket{0}$ and $\ket{1}$, which can be measured as output, the inverse quantum Fourier transform is applied. This makes it possible to calculate $\theta$ by measuring the phase estimation qubits.\footnote{QFT is not the only choice to recover the phase. Alternatives are listed, for example, in \cite{gomez2022survey}.} The process of measuring the phase estimation qubits forces each of these qubits to be either $\ket{0}$ or $\ket{1}$. This results in a binary string $\{b_{1}, b_{2}, \ldots, b_{n} \}$, where $b_k$ is the measured state of the $k$-th estimation qubit. Analogously with the Taylor approximation, the phase $\theta$ can then be computed from this binary string using the formula
\begin{equation}\label{Eq:mapping}
\theta = \sum_{i=1}^{n} \frac{b_{i}}{2^{i}}.
\end{equation}
Knowing $\theta$, the expected value $\mu$ can be calculated as specified in equation (\ref{mu_theta}).

However, it should be remembered that the quantum processes are inherently probabilistic. As a result, it is necessary to sample the quantum circuit numerous times. This results in a distribution of binary strings (as in Figure \ref{fig:example_normal} below). The most likely of these can then be used to calculate the estimate of the phase. This concludes the summary of the quantum amplitude estimation algorithm.

From equation (\ref{Eq:mapping}), it follows that that the error $\varepsilon_{\theta} = |\theta - \hat{\theta}|$ in estimating $\theta$ scales with the number of estimation qubits $n$ as $\varepsilon_{\theta} \propto \frac{1}{2^{n}}$.
Additionally, the number of applications $N$ of the unitary $\mathcal{Q}$ is approximately $2^{n}$. This comes from the fact that applying $\mathcal{Q}^k$ requires $k$ sequential applications of $\mathcal{Q}$ and that the sum of all applications of $Q$ for all estimation qubits is

\begin{equation}
    \sum_{i=0}^{n-1}2^i = 2^n-1.
\end{equation}
Thus, $N \propto \frac{1}{\varepsilon_{\theta}}$. It can also be shown that the error
$\varepsilon = |\mu - \hat{\mu}|$ in estimating the expectation value $\mu$ scales as $\varepsilon = \mathcal{O}(\varepsilon_{\theta})$, and therefore:
\begin{equation}
N = \mathcal{O}\left(\frac{1}{\varepsilon}\right).
\end{equation}
Hence, the QMC algorithm provides a quadratic speedup in the number, $N$, of applications of $\mathcal{Q}$ (a.k.a. “oracle calls").

With the QMC algorithm defined, it is natural to ask how the algorithm may be
practically implemented. To do so, we must decompose the probability-encoding unitary
$\mathcal{A}$, the random-variable encoding unitary $\mathcal{R}$, and the phase-encoding
unitary $\mathcal{Q}$ into elementary gates that are compatible with quantum hardware.

Practical applications of Monte Carlo estimation typically require evaluating expectation
values in a multidimensional
space, that is, as shown in equation \eqref{Eq:expectation_val} with $d > 1$. Here, it is often useful
to split up the register of discretization qubits into $d$ subregisters of $m$ qubits so
that each dimension in $\bm{x}$ is discretized into $2^{m}$ points. This approach is 
particularly fruitful when the joint probability distribution $p(\bm{x})$ is a product distribution so that each dimension of $\bm{x}$ is independent. A product distribution allows
$\mathcal{A}$ to be written as
\begin{equation}
\mathcal{A} = \bigotimes_{j=1}^{d} \mathcal{A}_{j},
\end{equation}
with each $\mathcal{A}_{j}$ applied independently to subregister $j$ and preparing the
corresponding marginal distribution. We focus on this use-case in the following.

\subsubsection*{Decomposing \texorpdfstring{$\mathcal{A}$}{A}\label{Sec:A_decomp}}

Encoding a probability distribution as a quantum state is in general non-trivial and is a
special case of state preparation. This can be performed using
exponentially-scaling circuits (\citealp{MottonenUCR, plesch2011quantum}), but such approaches
are insufficient for the QMC algorithm because the quadratic speedup will be lost. We hence
aim to identify efficient circuits that scale polynomially with the number of qubits in
the discretization register.

An efficient decomposition for $\mathcal{A}$, known as the Grover-Rudolph method (\citealp{GroverRudolph}), 
exists for probability distributions that are efficiently
integrable on a classical computer, such as log-concave distributions. However, 
there can be a significant overhead associated with calculating integrals in a preprocessing step. 
This is especially so when the integrals themselves need to be approximated with Monte Carlo estimation. 
As a result, it has been argued that the Grover-Rudolph method is insufficient for the QMC algorithm (\citealp{chakrabarti2021threshold, GroverRudolphProblem}).

Although alternative algorithms have been proposed by such authors as ~\cite{kaye2001quantum, kitaev2008wavefunction, kaneko2020quantum},
a major focus has been on optimizing variational quantum circuits to approximate a target 
distribution. The variational approach provides the flexibility to construct low-depth circuits
or increase the depth depending on the required accuracy. The work of \cite{Zoufal2019} proposes a hybrid
quantum-classical generative adversarial network and demonstrates its ability to prepare
canonical distributions, such as the log normal distribution. Here, we show how the probability
distribution of a variational circuit can be optimized directly using gradient-based
optimization.

Consider an $m$-qubit variational quantum circuit $U(\bm{\theta})$ composed of gates with 
trainable parameters $\bm{\theta}$ and applied to an input state $\ket{0}^{\otimes m}$.
When sampling in the computational basis, the circuit has a probability distribution
\begin{equation}
p(i, \bm{\theta}) = \left|\braket{i|U(\bm{\theta})|0}\right|^{2}.
\end{equation}
The objective is to minimize the distance between this distribution and a target
distribution $p_{\rm targ}(i)$. This can be achieved using the cost function
\begin{equation}\label{Eq:Cost}
C(\bm{\theta}) = \sum_{i} \left|p(i, \bm{\theta}) - p_{\rm targ}(i)\right|
\end{equation}
and finding
\begin{equation}
\bm{\theta}_{\rm opt} = {\rm argmin}_{\bm{\theta}} \,\,\,\, C(\bm{\theta}).
\end{equation}

Minimization of the cost function can be achieved using gradient-based optimizers by
calculating the gradient $\nabla_{\bm{\theta}}C(\bm{\theta})$, which is accessible on quantum
hardware using the finite-difference approximation or the parameter-shift rule for certain
types of circuit (\citealp{mitarai2018quantum, schuld2019evaluating}).
The resulting circuit $U(\bm{\theta}_{\rm opt})$ can then be used as an approximation to
$\mathcal{A}$ (or $\mathcal{A}_{j}$). ~\ref{app:a_train} shows how this method can be used to train a
variational circuit to reproduce the discretized normal distribution discussed
in section \ref{Sec:Example}.

\subsubsection*{Decomposing \texorpdfstring{$\mathcal{R}$}{R}}\label{Sec:R_decomp}

Similarly to the decomposition for $\mathcal{A}$, although approaches that scale exponentially
with the number of qubits in the discretization register exist for applying
$\mathcal{R}$ (\citealp{MottonenUCR}), doing so is insufficient to
preserve the quadratic speedup of the QMC algorithm. One commonly-used approach is to
discretize so that each dimension of $\bm{x}$ is represented as an $m$-bit fixed-point
number (\citealp{Derivatives, chakrabarti2021threshold}), for example, so that a positive $x$ can be written as
\begin{equation}
x = \sum_{i=1}^{m} 2^{m - i + k}b_{i},
\end{equation}
with a corresponding bitstring $\{b_{1}, b_{2}, \ldots, b_{m} \}$ and a fixed point
$k \in \mathbb{Z}$. An additional bit may be used to encode the sign of $x$.
We can then associate
the $m$-qubit basis state $\ket{b_{1} b_{2} \ldots b_{m}}$ with $x$.
With the random variable $f(\bm{x})$
written as a composition of such operations on $\bm{x}$, $\mathcal{R}$ can then be performed 
by applying those operations onto the corresponding registers.

Elementary arithmetic on a quantum circuit is a
well-established topic, and efficient
decompositions are available for common operations like addition
and multiplication.\footnote{See \cite{OptimizingArithmetic, chakrabarti2021threshold} for details.} Additional registers of calculation qubits may be required
to compute $f(\bm{x})$. The result must then be imprinted from a result
register onto the ancilla qubit. This can be achieved by first applying square-root and
arcsine operations and then performing a cascade of controlled-Y rotations.

Despite the existence of efficient decompositions, the elementary arithmetic approach to performing
$\mathcal{R}$ is challenging. Each operation can require a
non-negligible number of gates to enact as well as an additional register of qubits to store
the output, making the approach costly for implementation on hardware and simulators.
For example, adding two registers
of $m=5$ qubits using the approach of \cite{vedral1996quantum} requires $18$ Toffoli gates, $20$ CNOT gates, and an additional register
of qubits. This is an overhead that quickly adds up with multiple operations.
There are also often multiple decompositions provided in the literature with differing
advantages and disadvantages, and choosing which to adopt can be difficult.

Subsequent works have proposed alternative implementations of $\mathcal{R}$ with the objective
of avoiding quantum arithmetic as in \cite{Stamatopoulos_2020, herbert2021quantum}. In this
paper, we consider a simple approach outlined in  \cite{egger2020credit, Stamatopoulos_2020} for a
one-dimensional linear function $f(x)$. The details of this method can be found in ~\ref{app:linear_R}. Although
this setting is trivial to solve on a classical computer, it provides an entry point into
exploring the application of QMC to a relevant problem in economics and for benchmarking the
resource requirements, as we see in the following sections.

\subsubsection*{Decomposing \texorpdfstring{$\mathcal{Q}$}{Q}}\label{Sec:Q_decomp}

Recall that $\mathcal{Q} = (\mathcal{F}\mathcal{Z}\mathcal{F}^{\dagger} \mathcal{V})^{2}$
as given in equation (\ref{Eq:Q}). However, the QMC algorithm requires the ability to apply
$\mathcal{Q}$ with control from one of the phase estimation qubits. Due to the nature of
$\mathcal{Q}$, this can be achieved without the need for a controlled version of
$\mathcal{F}$, reducing the required depth of the algorithm. For further details, see \ref{app:Q_decomp}.

\subsection{Simple numerical example}\label{Sec:Example}

We now describe a simple Monte Carlo estimation problem and show how it can be solved using the
QMC algorithm with a simulator in PennyLane. Suppose we have a Gaussian probability distribution
with zero mean and unit variance:
\begin{equation}
p(x) = \frac{1}{\sqrt{2 \pi}} e^{- \frac{x^{2}}{2}},
\end{equation}
as well as a trigonometric random variable:
\begin{equation}
f(x) = \sin^2(x).
\end{equation}
The expectation value $\mu$ and variance $\sigma^{2}$ of $f(x)$ can be evaluated analytically
as
\begin{equation}
\mu = \frac{\sinh(1)}{e} \approx 0.432, \qquad
\sigma^{2} = \frac{\sinh^{2}(2)}{2 e^4} \approx 0.120.
\end{equation}

The first step is to discretize the problem. We set the grid variable to
\begin{equation}
\label{eq:disc_ex}
x_{i} = - x_{\rm max} + i \times \frac{2 x_{\rm max}}{2^{M} - 1},
\end{equation}
for $p$ and $f$ and (at the expense of abusing notation) define their values on the grid as
$$
p(i) = \frac{p(x_{i})}{\sum_{i \in X} p(x_{i})}, \notag
$$
$$
f(i) = \sin^{2}(x_{i}).
$$
~\ref{app:simple_example} shows how the problem can be discretized for $x_{\rm max} = \pi$ using $m=5$ qubits ($M = 32$) and $n=6$ phase estimation qubits. The QMC algorithm can be simulated in
PennyLane using the \texttt{QuantumMonteCarlo} template, as shown in the appendix.

\begin{figure}[t]
\includegraphics[width=0.65\textwidth]{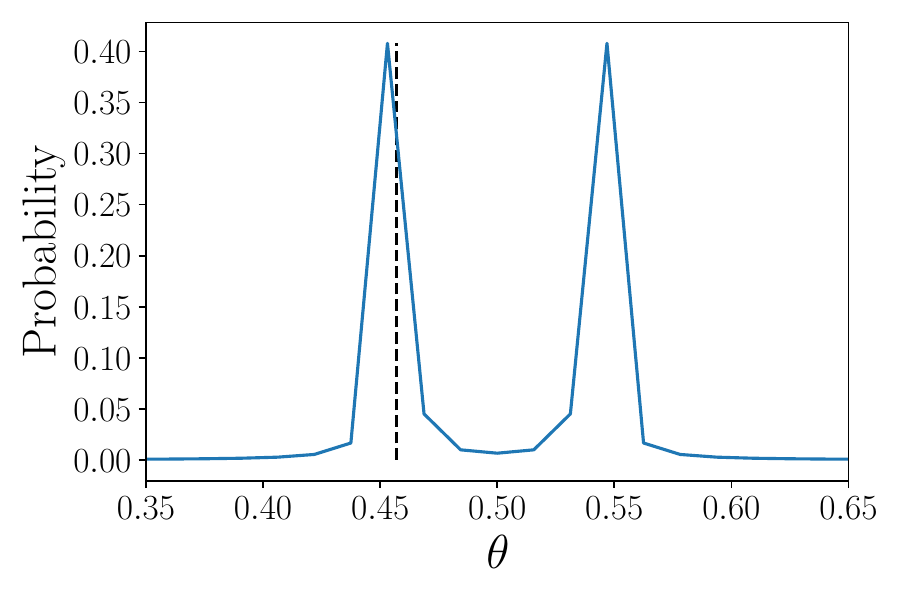}
\caption{Estimating $\theta$ using the QMC algorithm in PennyLane. The blue line indicates
the probability of estimating a given value of $\theta$, which is derived from the bitstring probabilities according to equation (\ref{Eq:mapping}). The dashed vertical line shows the theoretical value
of $\theta = 0.457$. The second peak can be removed by renormalizing \(f(i)\), as described in section \ref{ApplyingAmplitude}. }
\label{fig:example_normal}
\end{figure}

In \ref{app:simple_example}, we present the code that outputs the probability distribution of sampling the register of phase estimation
qubits in the standard basis. This distribution is plotted in Figure~\ref{fig:example_normal}. Different from classical estimations, the distribution of $\theta$ is not unimodal but bimodal and symmetric around $0.5$. These two possibilities result because we perform phase estimation with an input state $\ket{\chi}$
that is in an equal superposition of the eigenstates $\ket{\chi}^{\pm}$ with eigenvalues
$e^{\pm 2 \pi i \theta}$. If we expect $\theta > 0.5$, we focus on the right-hand side of the distribution; otherwise, if $\theta \leq 0.5$, we focus on the left-hand side. If this bimodal property creates many difficulties, the problem can be re-specified such that
$\theta$ is always larger than $0.5$ by suitably renormalizing $f(i)$. This results in a loss of accuracy, but this can be compensated for by adding another qubit to the phase estimation register.

\section{Two QMC Applications in Economics}\label{Sec:Apps}
In this section, we elaborate on two problems: stress testing of banks and deep learning applied to dynamic stochastic general equilibrium (DSGE) macroeconomic models. We first give a brief
introduction to each problem and then show how the QMC algorithm can be applied, presenting outcomes from small-scale instances of the problems.

\subsection{Stress testing of banks}\label{Sec:stress_testing}

We begin by performing a stylized macro-prudential stress test of banks given a dynamic stochastic stress scenario. Such stress tests are usually performed by a macroprudential regulator or a central bank to determine whether the financial industry can withstand large economic shocks and propagate the systemic risk to the real economy (see \citealp{adrian2020stress}, for a standard stress test protocol). The exercise typically begins with a specific narrative. It is then quantified with the time series projections of key macroeconomic and financial variables, such as inflation, GDP, interest rates, unemployment rate, etc. For simplicity, we will call such projections the macro scenario. The scenario is then used to quantify realized losses and expectations shocks that banks face on loans, securities, and funding exposures. We will refer to these inputs as the balance sheet shocks.

Monte Carlo simulations can be suitable for a stress test exercise for two main reasons. First, for a given time point and loss type, a stress designer may prefer to use a distribution of inputs rather than a single-value projection when creating the balance sheet shocks. In this paper, we focus on simulating probabilities of defaults of loans---a critical input to any stress test of banks. This means the macro scenario would be used for the projection of the distribution of defaults of loans rather than just their realized values. Multiplicity in the inputs allows for multiplicity of stress-test outputs, which enriches the interpretations of how economic and financial vulnerabilities may respond to the stress. In countries with only a few stress episodes in the past, such as Canada, simulations in the form of bootstrapping can be a natural way of accounting for the under-representation of high probabilities of default in the historical sample. Similarly, distribution can be introduced around the income projections of banks. For the results of a real stress-test application with bootstrapping of this kind, see \citet{mfraf-appl}.

The second reason why Monte Carlo simulations are suitable for stress tests is that they allow for the modeling of complex behaviors of banks observed in real financial markets. This is specifically important for the macro- rather than micro-prudential stress tests, with former ones focusing on the externalities that financial institutions create when responding to stress. Different central banks tend to include different behavioral responses of banks and other financial institutions in their stress tests (see \citealp{doyne}). The behavioral adjustments utilized the most are sales of securities for the purpose of maintaining capital or liquidity positions of banks. When additional financial mechanisms are added to the model, such as funding cost tightening and contagion of defaults, it often becomes difficult to find an explicit solution for the model. This happens because of the complex side-effects that the behaviors of financial institutions have on each other. While some behavioral stress-testing models provide explicit equilibrium solutions without the need to utilize Monte Carlo techniques, such as \cite{dbs}, the majority of models continue to rely on Monte Carlo methods for two reasons. First, development of stress-testing models with theory-powered predictions is time-consuming and requires precise calibration of parameters. Second, Monte Carlo techniques may still be beneficial in running various scenario inputs in parallel and performing sensitivity analysis (see \citealp{iori2012agent}, for other benefits and specifics). This highlights the relevance of our application for many central bank models.

To illustrate how QMC might be applied to stress testing, we calculate the expected capital losses of two banks over the span of two time periods of a stress scenario.\footnote{The current quantum algorithm can easily manage more banks and more periods, albeit at the cost of additional qubits.} Without loss of generality, assume that before stress is applied, bank $i$ holds assets $a^{all}_i$, out of which mortgages constitute $a^m_i$, business loans $a^b_i$, and securities $a^s_i$. The bank partially funds itself with equity $e_i$; the rest is funded with long-term debt.
\begin{figure}[t]
\includegraphics[width=1\textwidth]{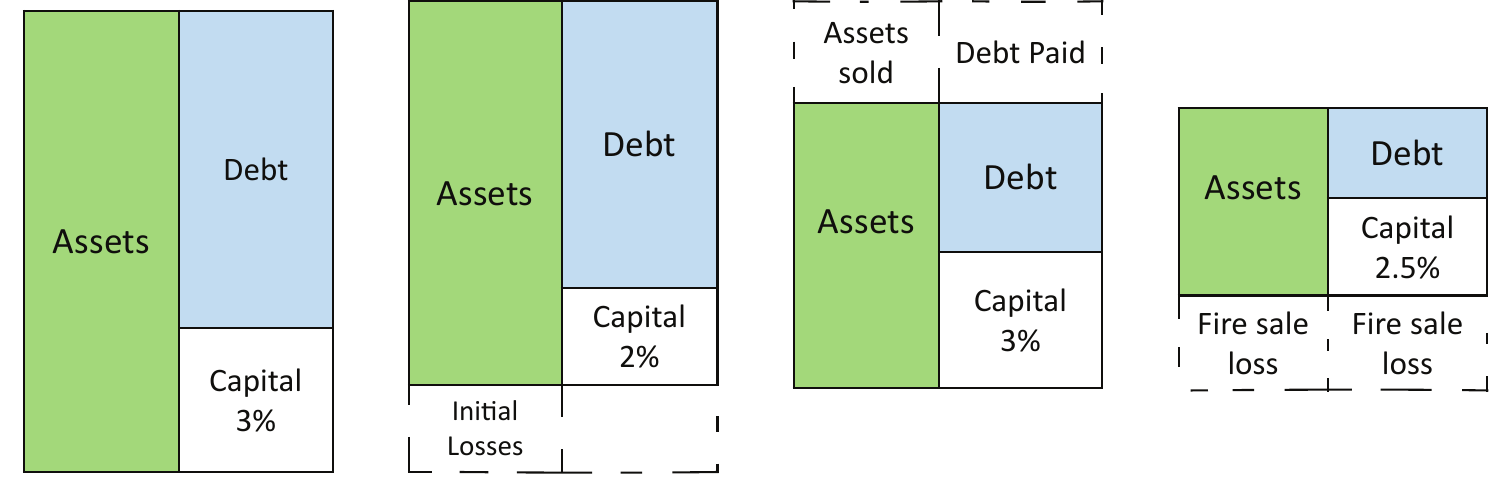}
\caption{Stylized evolution of the balance sheet of a bank following a financial shock. A sudden drop in the price of the assets reduces the bank's leverage ratio below the regulatory requirement. The bank sells assets to restore its leverage ratio. However, the rapid selling has an effect on prices of the assets, again reducing the leverage ratio and creating capital losses.}
\label{fig:bs}
\end{figure}
 
The exercise is performed period-by-period with the following sequence of events within each period $t$. At the beginning, bank $i$ faces credit losses 
\begin{equation}
loss^{cr}_{i,t} = a^b_{i} d^b_t+a^m_{i} d^m_t
\end{equation}
with default rates for mortgages and business loans
$$
 d^m_t = d^m_{t-1}(1+d_t),
$$
$$
d^b_t = d^b_{t-1}(1+d_t).
$$
Each period, more loans default at random, depending on the realization of the random variable $d_t$.\footnote{The difference between generating stochastic probabilities of default and credit losses is subtle. With the assumption that loss-given-default and utilization rate are both equal to one, the credit loss rate can be assumed to be equal to the probability of default.} 

We assume that pre-stress banks satisfy the regulatory leverage requirement to hold share $\beta$ of capital relative to total assets. After credit losses are applied, banks fall below their leverage requirement and are forced to adjust their balance sheets to come back to the required ratios. They do so differently in the near-term and in the long-term.\footnote{This split is an artificial modeling tool rather than a natural sequential order of actions.} 

In the short-term, each bank $i$ sells securities in the amount $\Delta a^s_{i,t}$, sufficient to restore the leverage ratio to threshold $\beta$. Because securities need to be sold quickly, this impacts the securities' market price and produces fire sales losses for the bank. For simplicity, we assume a linear price-response function:
\begin{equation}
\Delta p_t = - \alpha \sum_{i=1}^2 \Delta a^s_{i,t}. 
\end{equation}

Remaining securities on the balance sheets of all banks are also subject to re-evaluation due to mark-to-market accounting (see Figure \ref{fig:bs} for the balance sheet changes). As such, banks impose externalities on each other by selling securities simultaneously, and their capital ratios may fall below the regulatory requirement. This concludes the short-term response. 
In the long-term stage of each period, bank $i$ receives income and recapitalizes the equity to the initial level. The bank also restores the original portfolio of assets and liabilities. Capital injection may come from private sources or government support. Long-term adjustments of this kind are less typical for macro-prudential stress testing. Here, we introduce them for computational simplicity. Without going into the specifics of such costs, we assume that the regulator still cares about the capital losses of the first period. 

The changes between the initial and re-optimizing balance sheet can be found explicitly by doing some basic algebra. First, the amount of securities liquidated by each bank can be calculated as
$$
\Delta a^s_{i,t}=\frac{c_{i,t}-g_i c_{j,t}}{1-g_i g_j}
$$
with coefficients $c_{1,t}$ and $c_{2,t}$ being defined as linear function of initial credit losses,
\begin{equation}
c_{i,t}=\frac{e_i-(1-\beta) loss^{cr}_{i,t}-\beta a^{all}_i}{(1-\beta)\alpha a^s_i-\beta},
\end{equation}
and coefficients $g_1$ and $g_2$ being independent of the scenario,
\begin{equation}
g_i = \frac{\alpha(1-\beta) a^s_i}{(1-\beta)\alpha a^s_i-\beta}.
\end{equation}

The function of interest, total one-period loss as a fraction of industry assets, is thus linear in the default rates $d^m_t$ and $d^b_t:$
\begin{equation}
\frac{loss^{cr}_{1,t}+loss^{fs}_{2,t}+loss^{cr}_{1,t}+loss^{fs}_{2,t}}{a^{all}_1+a_2^{all}}.
\end{equation}

Therefore, computing the total expected system-wide loss over $T$ periods is equivalent to calculating the polynomial operator of random draws $d_t$:
\begin{equation}\label{Eq:RandomVariable}
E_0[loss^{\%}_T]=E_0[\gamma_1 \left(1+\gamma_0\frac{1-\beta}{\beta}\right)\sum_{t=1}^{T}\prod_{\tau=1}^{t}(1+d_{\tau})],
\end{equation}
where $\gamma_0$ captures fire sales impact,
$$
\gamma_0 = \alpha\frac{(1-g_1)(1-g_2)}{1-g_1g_2}(a^s_1+a^s_2),
$$
and $\gamma_1$ captures credit risk impact,
$$
\gamma_1 = \frac{(a^b_1+a^b_2)d^b_0+(a^m_1+a^m_2)d^m_0}{a^{all}_1+a^{all}_2}.
$$
\begin{table}[ht]
\centering
\caption{Stress test model calibration parameters.}
\begin{tabular}[t]{l|c|c|cc}
\hline
& &Bank 1&Bank 2&\\
\hline
Mortgages &$a^m$&50&30\\
Business Loans&$a^b$&50&70\\
Securities&$a^s$&50&50\\
Total assets&$a^{all}$&150&150\\
Equity&$e$&4.5&4.5\\
\hline
& &System-wide \\
\hline
Price sensitivity &$\alpha$& 0.0005\\
Leverage ratio&$\beta$ & 0.03\\
Benchmark mortgage loan loss & $d_0^m$ & 0.5\%\\
Benchmark business loan loss & $d_0^b$ & 1.5\%\\
Monte Carlo credit loss rate &$d$ & Beta(2,10)\\
\hline
Fire sales term &$\gamma_0$ & 0.0060 \\
Credit risk term &$\gamma_1$ & 0.0053
\label{tab:calib}
\end{tabular}
\end{table}%

We now discuss how this problem can be tackled with the QMC algorithm for two periods of stress with parameters as in Table \ref{tab:calib}. We specifically choose non-symmetric distribution of loss rates (Beta(2,10)) to account for fat tails typical for systemic risk events and to illustrate how the quantum methods can be applied to non-Gaussian distributions.

After substituting the parameters, equation (\ref{Eq:RandomVariable}) reduces to finding the expectation of 
\begin{equation}
\label{rv}    
0.0064\times(2+d_2)\times(1+d_1).
\end{equation}

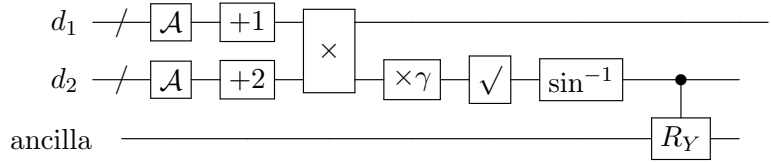
\begin{figure*}[t]
\mbox{
\Qcircuit @C=1em @R=0.5em {
\lstick{d_{1}} & {/} \qw & \gate{\mathcal{A}} & \gate{+1} & \multigate{1}{\rm \times}  & \qw  & \qw & \qw & \qw & \qw & \qw \\
\lstick{d_{2}} & {/} \qw & \gate{\mathcal{A}} & \gate{+2} & \ghost{\rm \times} & \gate{\times \gamma} & \gate{\sqrt{}} & \gate{\sin^{-1}} & \ctrl{1} & \qw \\
& \lstick{\rm ancilla\hspace{0.2cm}} & \qw & \qw &\qw &\qw & \qw & \qw & \gate{R_{Y}} & \qw 
% \gategroup{1}{3}{4}{3}{.7em}{--} \gategroup{1}{4}{4}{6}{.7em}{--}\gategroup{1}{7}{4}{9}{.7em}{--}\gategroup{1}{3}{4}{3}{.7em}{--}
}
}
\caption{A schematic circuit for the $\mathcal{F}$ unitary of the stress testing problem when estimating the total system-wide loss after $T=2$ periods. Note that the $d_{t}$ lines
represent registers of qubits and that $\gamma=\gamma_1 \left(1+\gamma_0\frac{1-\beta}{\beta}\right)$.}
\label{fig:stressTest_F_decomp}
\end{figure*}

For this we need to find $\mathcal{F}$
unitary that encodes the probability distribution---in our case $Beta(2,10)$---and the random variable---equation $(\ref{rv})$. As a first step, we see
that the problem can be written as a $T$-dimensional expectation value as in
equation~\eqref{Eq:expectation_val}, with a product distribution over the independent
$d_{t}$. We can hence discretize the input $d_{t}$ values each into
$M=2^{m}$ points so that there will be $T$ registers of $m$ qubits, and apply the
$\mathcal{A}$ unitary to each register to prepare the appropriate distribution
(see section \ref{Sec:A_decomp} for more details). A schematic for the procedure is provided in Figure~\ref{fig:stressTest_F_decomp} for $T=2$ periods.

Our next step is to provide access to the random variable given in equation~\eqref{Eq:RandomVariable} from which the
statistics are calculated. This can be achieved using the quantum arithmetic approach
discussed in \ref{Sec:R_decomp}, that is, so that each register represents a
fixed-point number and registers can be operated upon and combined to find the random variable. We first adjust the
$\mathcal{A}$ unitaries to shift the mean by $1$ or $2$. We can then apply a multiplication operation that multiplies the two registers, as shown in Figure~\ref{fig:stressTest_F_decomp}. Note that quantum
arithmetic operations may require additional registers of calculation qubits due to the
reversible nature of quantum computing. These registers are omitted here, and we assume that the output register of the calculation is situated on the bottom wire in
the diagram.

To finish calculating the random variable, we apply operations that multiply by a constant. Finally, the square root and arcsine operations are applied and the result is
imprinted onto the ancilla qubit using a controlled-Y cascade. The controlled-Y cascade places the ancilla into the state $\cos{(i)}\ket{0}+\sin{(i)}\ket{1}$, whereas the goal is to place the ancilla into the state $\sqrt{1-i}\ket{0}+\sqrt{i}\ket{1}$. This is why the cascade is preceded by the square-root and the arcsine operations. The work of~\cite{chakrabarti2021threshold} provides a summary
of decompositions for typical operations, including multiply, square root, and arcsine.% One
%approach to calculate the minimum is to use %the identity ${\rm min}\{a, b\} = (a + b - %|a - b|) / 2$.

\begin{figure}[t]
\includegraphics[width=0.65\textwidth]{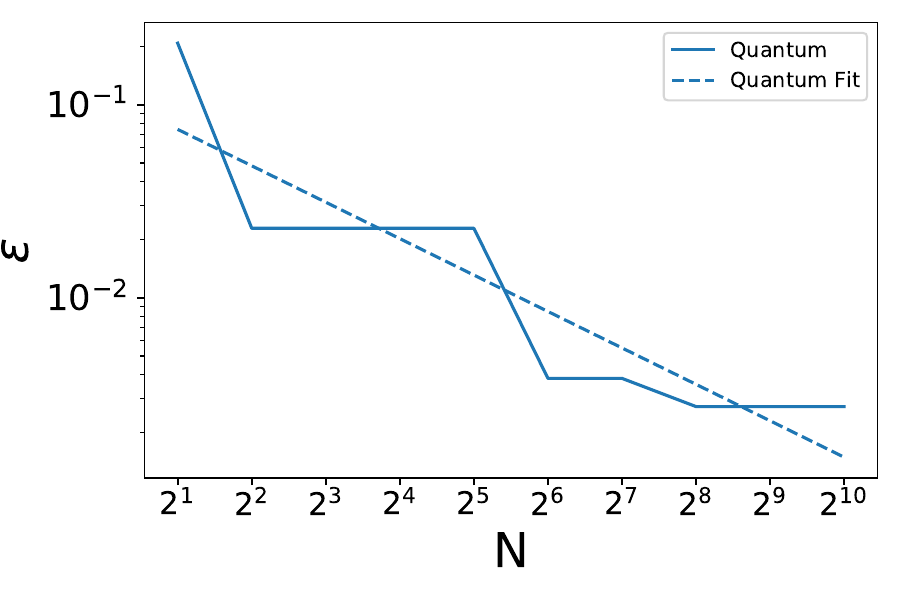}
\caption{The error of estimating the total loss of the banking industry as a fraction of the actual (theoretical) value versus the number of oracle applications within the QMC algorithm in log-log scale. Notice that for $N>10^6$, discretization error begins to dominate. Therefore, to continue seeing the reduction of the computational error, $\varepsilon$, the number of discretization qubits per variable would need to be increased.}
\label{fig:stress_testing}
\end{figure}

With $\mathcal{F}$ defined, the QMC algorithm can then be carried
out by constructing $\mathcal{Q}$ according to the decomposition in \ref{Sec:Q_decomp}. The results of this are shown in Figure \ref{fig:stress_testing}. 

With $10$ estimation qubits, QMC finds the total banking industry loss over two periods as a fraction of total assets to be $1.622\%$, as compared to the theoretical value of $1.618\%$. This corresponds to the error of $0.0027$ as a fraction of the theoretical value. However, even just $2$ estimation qubits give a good approximation, resulting in the fractional error of $0.023$.

We have hence built up the protocol for an example instance of stress testing using quantum
computation. Notably, we find that the total number of (logical) qubits required for this problem is manageable in theory. However, in practice, the current quantum hardware is noisy. The noise begins to dominate the results after around $100$ gates (\citealp{lubinski2021application}). Meanwhile, our problem requires at least $10^4-10^5$ gates, as we show in the next section.

\subsection{Deep learning solutions of DSGE models}\label{sec:growth}

Another promising application of the QMC algorithm within economics is solving DSGE models with deep learning as in \cite{maliar2021deep}, \cite{fernandez2019financial}, and \cite{azinovic2019deep}.\footnote{For an introduction into deep learning, see \cite{goodfellow2016deep}.} DSGE models are used in macroeconomics to study relationships between aggregated economic variables such as inflation, gross domestic product, consumption, and capital goods, often in the context of government policies. Standard methods of solving such models suffer from the curse of dimensionality: computational time grows exponentially with the number of state variables and can become unmanageable beyond a small number of variables. By contrast, methods based on deep neural networks can break this curse (\citealp{bach2017breaking}) and are now being applied to solve large economic models (see, for instance, \citealp{lepetyuk2020us}).

However, the deep learning approach has its own shortcoming. It relies on Monte Carlo sampling to average over all future outcomes (shocks) to make proper adjustments of neural network parameters. This results in a slow convergence rate of the computational error (this is demonstrated in section \ref{sec:cons_saving_model}). Thus, even though the deep learning method enables one to solve models with thousands of variables, the resulting solutions might be less accurate in practice than what computational economists may be used to with traditional methods.

This is where QMC could be helpful. By speeding up averaging over future outcomes, QMC can increase the rate of convergence of the deep learning method, thus increasing its accuracy in practice.

\subsubsection{Neoclassical investment model}\label{sec:neoclass}

To get an understanding of when quantum Monte Carlo might offer an advantage over the standard approach for this application, we consider a simple neoclassical stochastic investment model (\citealp{maliar2015merging}, \citealp{stokey1989recursive}). In this model, the representative agent starts each period with capital, \(k\), and current level of productivity, \(z\). The agent's task is to decide how much to consume right away, \(c\), and how much capital to leave for next period, \(k'\). The capital will be invested into production, which will deliver future payoff according to the production function $f(k)$. Part of the capital $\delta$ will depreciate, while the rest can be reused. Output of production will also vary due to the randomness in the productivity of the agent. The future level of productivity, \(z'\), depends on the current level of productivity, $z$. For simplicity, assume that agents value their current consumption according to the utility function \(u(c)\) and the flow of their future consumption according to the discounting factor $\beta$. This means that the agent computes the value function from the following Bellman equation:
\begin{equation} \label{V}
    V(k,z)=\max_{c,k'}u(c)+\beta E[V(k',z')]
\end{equation}
subject to:
\begin{gather}
    c+k' = zf(k)+(1-\delta)k, \label{c1}\\
    \ln{z'}=\rho\ln{z}+\sigma\epsilon,\quad \epsilon \sim \mathcal{N}(0,1),\label{c2}
\end{gather}
where $u$ and $f$ are strictly increasing, continuously differentiable, and concave; \(\beta \in (0,1)\);
\(\delta \in (0,1]\); \(\rho \in (-1,1)\); and \(\sigma \geq 0\). Under these assumptions, the problem has a unique solution (\citealp{stokey1989recursive}).

The general algorithm for solving this problem with deep learning proceeds as follows:
\begin{enumerate}
    \item Interpolate \(V(k,z)\) by a neural network with learnable parameters \(\{s_i\}\). In the current work, the interpolation part is done on a classical computer. However, in the future, a quantum variational algorithm (\citealp{cerezo2021variational}) could be a viable alternative to the classical neural network.
    \item Draw \(N_s\) random samples of 
    \begin{equation}
    \label{z1z2}
    (k,z,z'_1,z'_2),
    \end{equation}
    where \(k \sim \mathcal{U}_{[k_{min},k_{max}]}\), \(z \sim \mathcal{N}(1,\sigma^2)\), and $z'_1$ and $z'_2$ are two possible realizations of the random process (\ref{c2}).
    
    \item For each sample, use the first-order conditions and the envelope condition to compute consumption \(c\):
    \begin{gather}
        u_c(c)=\frac{V_k(k,z)}{zf_k(k)+1-\delta}\\
        \Rightarrow c=u_c^{-1}\left(\frac{V_k(k,z)}{zf_k(k)+1-\delta}\right),
    \end{gather}
    where \(u_c(c)=\frac{\partial u(c)}{\partial c}\), \(V_k(k,z)=\frac{\partial V(k,z)}{\partial k}\), \(f_k(k)=\frac{\partial f(k)}{\partial k}\).
    \item Solve for \(k'\) using the budget constraint (equation \eqref{c1}).
    \item Compute the Bellman error for each next-period productivity \(z'_i\):
    \begin{equation}
        B_E(k,z,z'_i)=u(c)+\beta V(k',z'_i)-V(k,z).
    \end{equation}
    \item Calculate the mean squared error (MSE) over the entire Monte Carlo sample, defined as
    \begin{equation} \label{bellman}
        \frac{1}{N_s}\left|\sum_j^{N_s} B_E(k_j,z_j,z'_{j_1})B_E(k_j,z_j,z'_{j_2})\right|.
    \end{equation}
    \item If the MSE is less than the predefined threshold, stop the computation. Otherwise, update (via backpropagation or parameter-shift rule (\citealp{crooks2019gradients})) parameters \(\{s_i\}\) to minimize the MSE.
\end{enumerate}

Within this algorithm, QMC can be used to estimate the MSE in expression (\ref{bellman}). Then, using tools such as the PennyLane library, a classical computer can be used to calculate gradients and update the learnable neural network parameters.

Although expression (\ref{bellman}) can be used with QMC directly (and we do so in the next section), realizing the quantum advantage in practice requires reducing the depth of the quantum circuit as much as possible. Toward this end, we make several simplifications and then expand the problem in terms of quantum gates. This makes it possible for us to find where (for various quantum gate times) QMC gains an advantage over classical MC.

First, we make the standard choice of the utility and production functions:
\begin{equation}
%\begin{align}
    u(c) =
    \begin{cases}
        \frac{c^{1-\theta}-1}{1-\theta} & \theta \neq 1 \\
        \ln{c} & \theta=1
    \end{cases},
%\end{align}
\end{equation}
\begin{equation}
%\begin{align}
f(k) = k^\alpha.
%\end{align}
\end{equation}
Furthermore, we make a simplifying choice of \(\theta=1\), and \(\delta = 1\). With this simplification, the problem has an analytical solution:
\begin{equation} \label{loglinear}
    V(k,z)=R\ln{k}+S\ln{z}+Q,
\end{equation}
where \(R=\frac{\alpha}{1-\alpha\beta}\), \(S=\frac{1+\beta R}{1-\beta\rho}\), and \(Q=\frac{\ln(1-\alpha\beta)}{1-\beta}+\frac{\beta R\ln(\alpha\beta)}{1-\beta}\).
Since this is a linear equation (in the variables $\ln{k}$ and $\ln{z}$), a neural network composed of a single layer with weights \(s_1\), \(s_2\), and bias \(s_0\) (which is equivalent to a linear function) has enough model capacity to learn this analytical solution exactly. This makes it possible to easily express the Bellman error in terms of the input random variables \(k,z,z'_1,z'_2\). Leaving the solution of the full problem for later work when quantum computation technology is more advanced, we take \(k\) and \(z\) to be constant within each random Monte Carlo sample. Moreover, we replace expression \eqref{bellman} with the following loss function:
\begin{equation}\label{selectedloss}
    \frac{1}{N_s}\left|\sum_j^{N_s}B_E(k_j,z_j,z'_j)\right|.
\end{equation}
This loss function has a minimum at the same place as the original MSE loss for this particular problem, allowing us to keep only a single \(z'\) random variable. As a final simplification, since \(z'\approx 1\), we assume \(\ln{z'} \approx z'-1\).

Interpolating the value function in terms of the chosen neural network architecture,
\begin{equation}
    V(k,z)=s_1\ln{k}+s_2\ln{z}+s_0,
\end{equation}
and with the simplifications above, the chosen loss function \eqref{selectedloss} reduces to
\begin{equation}%\min_{s_0,s_1,s_2}
\label{optimalSol}
    \left|\frac{1}{N_s}\sum_{j=1}^{N_s}(C_1+C_2 z_j')\right|,
\end{equation}
where
\begin{align}
    C_1&=(s_1-\alpha(1+\beta s_1))\ln(k)+ (s_2-(1+\beta s_1))\ln(z)& 
\end{align}
    $$
    +(1-\beta)s_0 
    -\ln\left(\frac{\alpha}{s_1}\right)-\beta s_1\ln\left(1-\frac{\alpha}{s_1}\right) +\beta s_2
$$
and
\begin{align}
C_2& =-\beta s_2. &&&&&&&&&&&&&
\end{align}

The simplified neoclassical model is therefore expressed as the expectation value of a linear random variable with respect to the normal distribution. This provides us with a simple, application-relevant use case to investigate and benchmark the QMC approach.

Our first step is to set the values $C_{1} = 30$ and $C_{2} = -29$. The exact choice of these constants is not important, but we pick these values since they are close to the optimal values (as given in equation (\ref{optimalSol})) and could realistically occur during the training of the neural network. The distribution of the random variable $z'$ is normal with mean $\mu = 1$, and we choose $\sigma=0.02$. We then implement the QMC algorithm as outlined in section \ref{Sec:R_decomp}. The details of our QMC implementation for this problem can be found in \ref{app:QCMNeoclassical}.

\begin{figure}[t]
\includegraphics[width=0.65\textwidth]{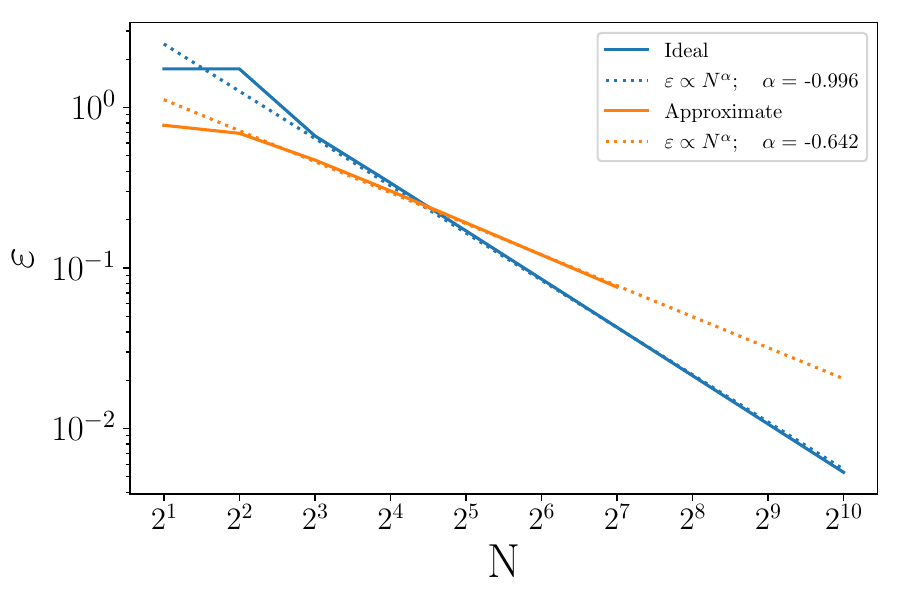}
\caption{Finding the estimation error $\varepsilon$ when using
the QMC algorithm with a time complexity of $N$ in log-log scale. The blue line 
shows the ideal case when the unitary $\mathcal{R}$ can be realized
exactly, while the orange line illustrates the use of the linear
approximation outlined in ~\cite{egger2020credit, Stamatopoulos_2020}. The dotted lines show a linear fit in
log-log scale.}
\label{fig:qmc_scaling}
\end{figure}

The QMC algorithm provides an estimate of $\hat{\theta}$ that can
be converted into an estimate of $\hat{\mu}$ using
equation~\eqref{Eq:Theta} and accounting for the normalization of $f(x)$. Figure~\ref{fig:qmc_scaling} illustrates how the error
$\varepsilon = |\mu - \hat{\mu}|$ scales with the number $N$ of
applications of $\mathcal{Q}$.

We plot the error in two cases:
(orange line) when using the linear approximation detailed in \ref{app:QCMNeoclassical}, and
(blue line) when using an exact simulation of $\mathcal{F}$.
The linear approximation has a worse theoretical speedup, but does not require as many quantum resources and so can be realized sooner in practice. The plot shows different scaling in each case, as expected. The linear
fit in log-log scale gives a slope of $\alpha = -0.642$ 
in the approximate case (theoretical value: $-\frac{2}{3}$) and a slope of $\alpha = -0.996$ in the
exact case (theoretical value: $-1$), both of which are better than the scaling of classical Monte Carlo estimation, $\alpha = -\frac{1}{2}$.

\subsubsection{Consumption-saving problem with occasionally binding borrowing constraint and exogenous stochastic shocks}\label{sec:cons_saving_model}

To illustrate that the quantum speedup persists for more realistic models, we enhance the previously considered DSGE model in line with \cite{maliar2021deep}, whose computer code is available at \cite{maliardeep_code}:
\begin{gather*}
\underset{\left\{ c_{t},w_{t+1}\right\}_{t=0}^{\infty }}{\max }E_{0}\left[
\sum_{t=0}^{\infty }\exp (\delta_{t})\beta ^{t}u\left( {c_{t}}\right)\right]  \\
\text{s.t. }w_{t+1}=\left( w_{t}-c_{t}\right) \overline{r}\exp (r_{t+1})+\exp
(y_{t+1}), \\
c_{t}\leq w_{t}.
\end{gather*}

The dynamics of the consumption-savings decision is now tracked using consumption $c_t$ and cash-on-hand $w_{t}$ at the beginning-of-period. Savings are growing with a (gross)
constant interest rate $\overline{r}\in \left( 0,\frac{1}{\beta }\right)$.

 In addition, the total income consists of random (transitory) income component $p_{t}$ and permanent income component $q_t$:
 $$
y_{t}=\exp(p_{t})\exp(q_{t}).
 $$
 Therefore, there are four different exogenous
state variables, namely shocks to the interest rate $r_{t}$, discount factor $\delta_t$, and income components $q_{t}$ and $p_{t}$. All
exogenous variables  follow AR(1) processes with error terms having standard normal distribution: 

$$
{y_{t+1}} =\rho_{y} {y_{t}}+\sigma_{y}\epsilon^{y}_{t+1}
$$
$$
{p_{t+1}} =\rho_{p}{p_{t}}+\sigma_{p}\epsilon^{p}_{t+1}
$$
$$
{r_{t+1}} =\rho_{r}{r_{t}}+\sigma_{r}\epsilon^{r}_{t+1}
$$
$$
{\delta_{t+1}} =\rho_{\delta }{\delta_{t}}+\sigma_{\delta }\epsilon^{\delta}_{t+1}.
%\end{eqnarray}
$$
The utility function is constant relative risk aversion (CRRA): $u\left( {c_{t}}\right) =\frac{c_{t}^{1-\gamma }-1}{1-\gamma }$.

The classical algorithm proceeds by taking $N$ random draws of all the endogenous and exogenous variables. Each of the four exogenous variables corresponds to two random variables in the algorithm because we draw two realizations of future shocks (see expression (\ref{z1z2})). Together with the five endogenous variables, this results in thirteen random variables sampled $N$ times each. These variables are inputted into the neural network (with three $32$ by $32$ fully connected linear layers joined by ReLU activation functions) to calculate the Bellman error, analogously to how this was done in the previous section.

\begin{figure}[t]
\includegraphics[width=0.65\textwidth]{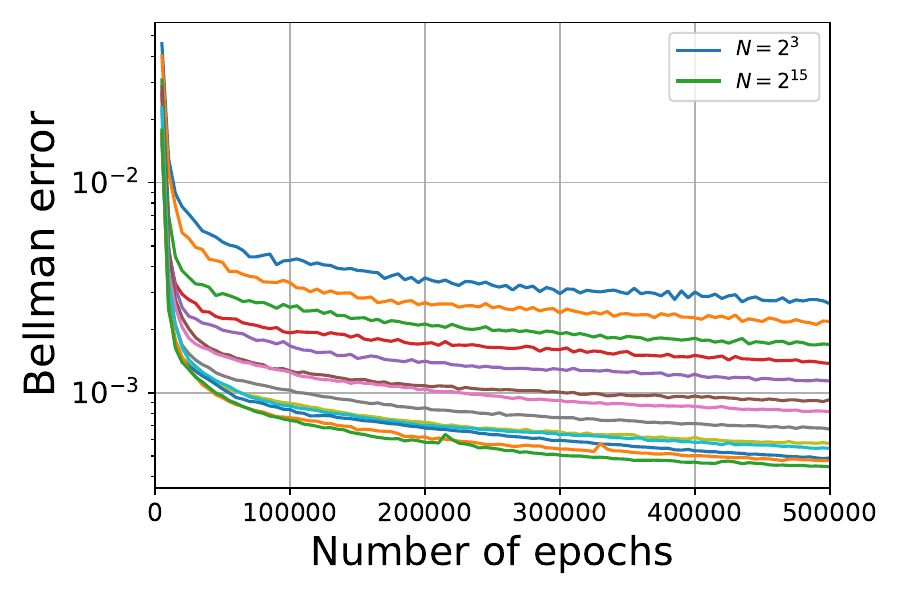}
\caption{Bellman error vs. number of epochs for various number of random draws, $N$. $N$ varies from $2^{3}$ to $2^{15}$. The Bellman errors plotted are averages across $5000$ epochs.}
\label{fig:cons_save_epochs}
\end{figure}

Figure \ref{fig:cons_save_epochs} shows how the Bellman error changes during the first $500,000$ epochs (computational steps in which the neural network parameters are updated) for different number of random draws, $N$. Table \ref{cons_save_table} shows how the computational time changes with the number of random draws, $N$. The computational error scales with regard to $N$ with the exponent $-0.22$ (this exponent appears to be largely independent of the number of epochs taken). The error also scales approximately with the exponent $-0.26$ in terms of the number of epochs, but epochs are sequential and, unlike Monte Carlo draws, cannot be parallelized. Thus, to reach the Bellman error of $10^{-5}$ classically after $500,000$ epochs, over $10^{11}$ MC draws every epoch are necessary. To reach the computational error of $10^{-8}$, this number rises to over $10^{25}$. This illustrates the difficulty of achieving high accuracy with the deep learning method.

\begin{table*}[!tb]
\[\begin{array}{c|c|c}
\hline
N & t_{500,000} (s) & \varepsilon_{500,000} \\ \hline 
8 & 522 & 2.655 \times 10^{-3}  \\
16 & 524 & 2.189 \times 10^{-3}  \\
32 & 534 & 1.698 \times 10^{-3}  \\
64 & 525 & 1.379 \times 10^{-3}  \\
128 & 564 & 1.139 \times 10^{-3}  \\
256 & 633 & 0.928 \times 10^{-3}  \\
512 & 755 & 0.813 \times 10^{-3}  \\
1024 & 1133 & 0.671 \times 10^{-3}  \\
2048 & 1716 & 0.572 \times 10^{-3}  \\
4096 & 4097 & 0.544 \times 10^{-3}  \\
8192 & 9579 & 0.490 \times 10^{-3}  \\
16384 & 17523 & 0.476 \times 10^{-3}  \\
32768 & 32679 & 0.445 \times 10^{-3} 
\end{array}\]
\caption{For the consumption-saving model, this table lists the computational time (seconds) and the final error after 500,000 training epochs as the number, $N$, of MC draws per epoch is changed.}
\label{cons_save_table}
\end{table*}

\begin{figure}[t]
\includegraphics[width=0.65\textwidth]{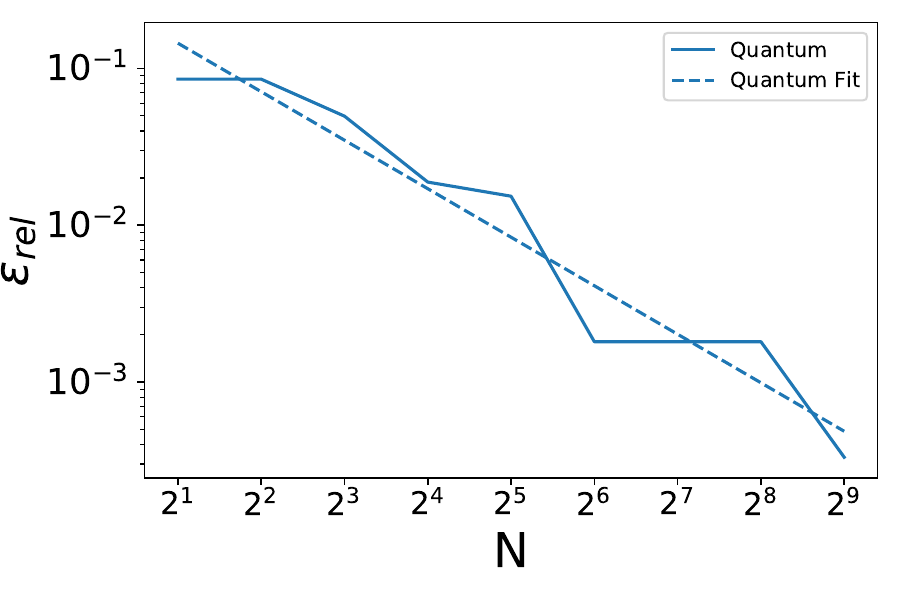}
\caption{The precision of solution (MSE in expression (\ref{bellman})) vs. the number of quantum oracle calls in log-log scale for the consumption-saving model with two random variables. The scaling exponent of the fitted line is $-1.0283$.}
\label{fig:cons_save_2vars}
\end{figure}

\begin{figure}[t]
\includegraphics[width=0.65\textwidth]{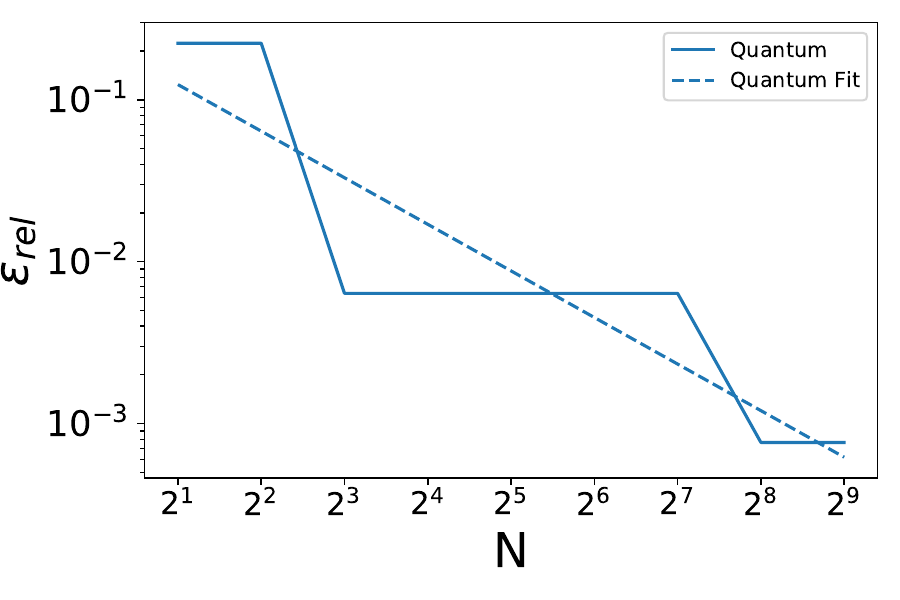}
\caption{The MSE in expression (\ref{bellman}) vs. the number of quantum oracle calls for the consumption-saving model with three random variables. The scaling exponent is $-0.9557$.}
\label{fig:cons_save_3vars}
\end{figure}

This is one example where QMC may be helpful in the future. The quadratic speedup of QMC persists even when multiple random variables are present (\citealp{cornelissen2021quantum}). We have demonstrated this once already in section \ref{Sec:stress_testing} when solving the stress testing scenario, which had two random variables. We further demonstrate this in Figures \ref{fig:cons_save_2vars} and \ref{fig:cons_save_3vars} for the consumption-saving problem.

Figure \ref{fig:cons_save_2vars} shows how accurately QMC computes the MSE in expression (\ref{bellman}) during one epoch for a model that has two random variables: two endogenous variables, $w_t$ (cash-on-hand) and $r_t$ (interest rate) and no shocks $r_{t+1}$ since these are separate random variables. As seen in the figure, QMC follows the expected scaling law, with an exponent equal to $-1.0283$. Similarly, Figure \ref{fig:cons_save_3vars} shows the scaling law when the model has three random variables: the interest rate, $r_t$ and both of its future shocks, $r_{t+1,1}$ and $r_{t+1,2}$. As before, the scaling law is as expected, with the exponent equal to $-0.9557$.

We stopped at three variables, because this is what can be easily simulated on a classical computer. Going to additional variables would already require a high-performance computing solution or, of course, a fault-tolerant quantum computer. In fact, even three variables is already cumbersome. To be able to simulate the circuit, the number of discretization qubits has to be reduced from $5$ to $3$ per variable. Regardless, as long as the additional variables are scaled appropriately (\citealp{cornelissen2021quantum}), additional variables do not present an issue to QMC's quadratic speedup.

However, there is a difference between a theoretical speedup of quantum oracle calls vs. number of random classical samples and between a practical speedup in computational time. That is, the proportionality constant might be so large for the quantum solution that any kind of speed gains can never be achieved in practice. For the simple neoclassical model, we can investigate this matter directly. This is the subject of section \ref{Sec:Bench}. For the more realistic consumption-saving model, the existence of the speedup in practice will depend on how efficiently the model is encoded onto the quantum computer. We will discuss how this might be realized in section \ref{sec:extensions}.

\section{Benchmarking of Quantum and Classical Methods}\label{Sec:Bench}

We have established that the QMC algorithm has a speedup
in terms of oracle calls $N$ when compared to its classical
counterpart. However, in the QMC algorithm, $N$ is the number
of applications of $\mathcal{Q}$, while in classical Monte
Carlo $N$ is simply the number of samples drawn. To provide a fair
comparison between the two approaches requires working out the real
time requirements for both implementing $\mathcal{Q}$ and for
drawing a sample from a classical random number generator. In this
section we provide a time-based benchmark of QMC using the setting
of the simple neoclassical macroeconomic deep learning problem from the previous section.
This should build intuition about what should be possible to achieve with QMC in the future.

\begin{figure}[t]
\includegraphics[width=0.65\textwidth]{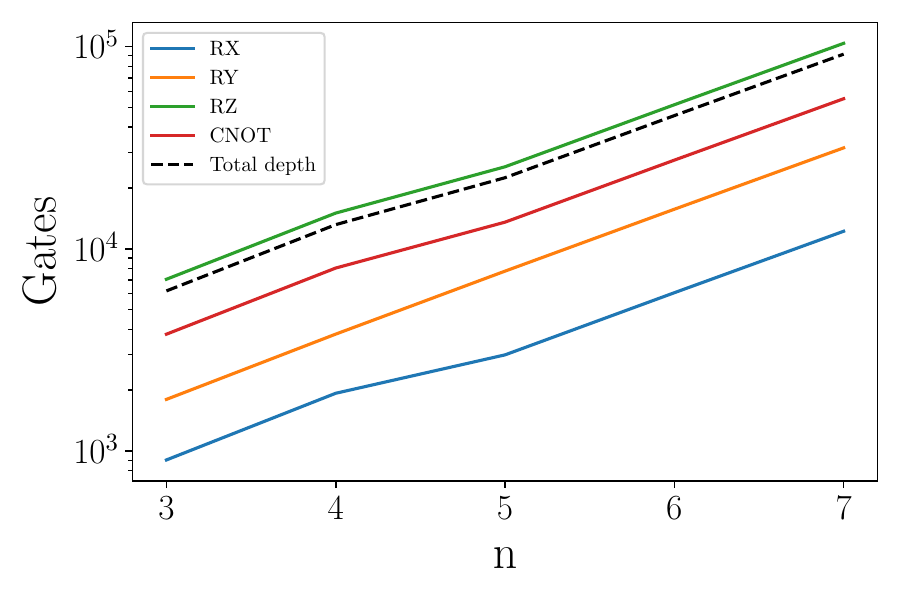}
\caption{Counting the number of elementary gates required to
implement the QMC algorithm (approximate $\mathcal{F}$) for varying numbers of phase estimation
qubits $n$. The circuit depth is also drawn as a dashed line.}
\label{fig:qmc_gates}
\end{figure}

Our first step is to understand the resource requirements of
the QMC algorithm for varying numbers $n$ of phase estimation 
qubits. Using the decompositions discussed in section \ref{Sec:QMC}, it is
possible to break down the QMC algorithm into elementary gates that
are compatible with hardware implementation. The code for doing so can be found in \ref{app:QCMNeoclassical}. We choose a gate
set composed of the
single qubit rotations $R_{X}$, $R_{Y}$, and $R_{Z}$, as well as
the two-qubit CNOT gate. Figure~\ref{fig:qmc_gates} provides a
gate count for implementing the QMC algorithm for the macroeconomic problem using the linear approximation with $m=5$
discretization qubits and a range of phase estimation qubits $n$. We focus on the approximate case because it requires significantly fewer quantum resources than the exact case and so is likely to be realized in practice first.
As expected, the gate counts scale
exponentially since the number of applications of $\mathcal{Q}$
is $2^{n}$.

As well as counting individual gates, we can also calculate the
depth of the QMC algorithm, that is, the longest sequential path
of gates through the circuit. The depth is also shown in Figure~\ref{fig:qmc_gates} as the dashed line and can be extrapolated
using a log-linear fit to arbitrary $n$. 

From the depth $d$, we can
calculate the algorithm time $T_{\rm tot}$ by assuming each gate 
can be run in time at most $t$ and finding $T_{\rm tot} = td$. We consider gate times ranging from $1 \,\,{\rm ns}$ to $100 \,\,{\rm \mu s}$. This in accordance with the various gate times already in existence for various quantum computing hardware, as discussed in section \ref{Sec:q_hardware}, though we believe it entirely possible that gate times could become even shorter as technology develops.

\begin{figure}[t!]
\includegraphics[width=0.65\textwidth]{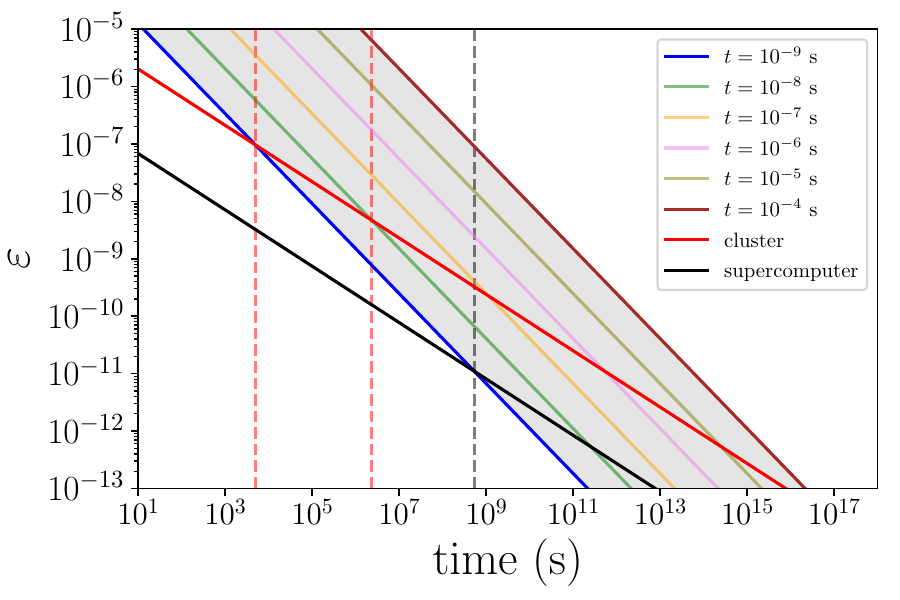}
\caption{Providing a time-based comparison for the
macroeconomic model discussed in section \ref{sec:neoclass} between the QMC algorithm (with a linear approximation of the $\mathcal{R}$ unitary) for a variety of gate
times $t$ (shaded area) and between two classical HPC systems: a supercomputer (black line) and strong HPC cluster (red line), as defined in the text. The estimation
error $\varepsilon$ is plotted against the algorithm time $T_{\rm tot}$. These quantities
are extrapolated for the QMC algorithm using the linear approximation of $\mathcal{R}$ for a range of phase
estimation qubits $n$. Dashed vertical lines illustrate where the quantum and classical algorithms coincide at certain points.}
\label{fig:qmc_time}
\end{figure}

To obtain the plot of the error $\varepsilon$ vs total time $T_{\rm tot}$, recall that the fit in Figure~\ref{fig:qmc_scaling} can be used to
find the error scaling of the approximate approach for a given number of estimation qubits $n$. The number of estimation qubits is related to the circuit depth (Figure \ref{fig:qmc_gates}), which is related to the total time. This leads to Figure \ref{fig:qmc_time}, where we plot $\varepsilon$ against $T_{\rm tot}$ for a variety of gate times. This figure
provides the basis for a time-based comparison of the QMC
algorithm to standard Monte Carlo estimation.

To generate classical Monte Carlo results in Figure \ref{fig:qmc_time}, we find the mean average error of estimating the same
expectation value as the one obtained using a range of random draw numbers $N$.
We also record
the time taken for each $N$ and extrapolate using a linear fit in
log-log scale. Furthermore, since the quantum computer would be competing against parallel computing systems rather than just a single CPU, we extrapolate the classical result to two different situations. As an order-of-magnitude estimate, we define a strong HPC cluster as a system that provides a $1,000$-fold speedup over a single CPU core. Likewise, we define a supercomputer as providing a speedup of $1,000,000$-fold.

Figure~\ref{fig:qmc_time} illustrates the expected speedup (or lack thereof)
provided by the QMC algorithm (with a linear approximation of the $\mathcal{R}$ unitary) due to the steeper slope of
the extrapolated line ($-\frac{2}{3}$ for quantum versus $-\frac{1}{2}$ for classical). The choice of individual gate times $t$
sets the offset height of each line for QMC, which has the
practical implication of determining when the QMC algorithm
becomes preferential to its classical counterpart. A smaller $t$
results in the QMC line crossing the classical line at a lower
value of total algorithm runtime $T_{\rm tot}$. Some crossover
points are shown as dashed vertical lines in the figure.

The figure shows that a classical supercomputer is likely to remain quicker than (approximate) QMC for many years to come. However, if gate times are lower than $10$ns, QMC can outperform strong computational clusters that an average researcher could have access to. For example, assuming the gate time of $1$ns, the quantum computer reaches the computational error of $10^{-8}$ $5.6$ times faster than the HPC cluster defined above. If the desired computational error were lower, the quantum advantage would increase.

\begin{figure}[t!]
\includegraphics[width=0.65\textwidth]{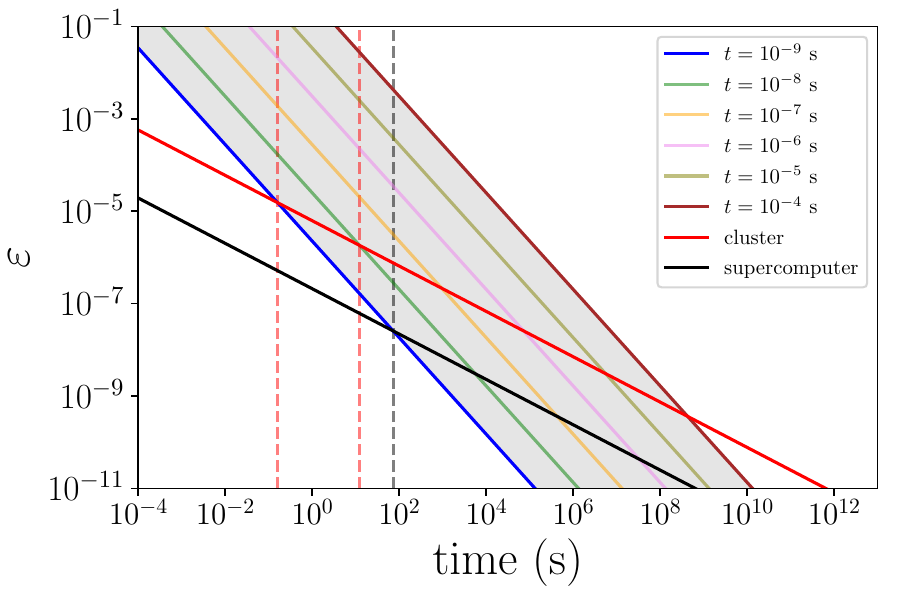}
\caption{A hypothetical extension of Figure \ref{fig:qmc_time}, such that the depths of the QMC algorithm remain unchanged but the QMC algorithm has the ideal scaling exponent (equal to $-1$).}
\label{fig:qmc_time_hypothetical}
\end{figure}

To illustrate what might be possible eventually, Figure \ref{fig:qmc_time_hypothetical} considers a hypothetical scenario that the QMC algorithm has the ideal scaling law (scaling exponent equal to $-1$) but retains the depth of the approximate QMC algorithm (see Figure \ref{fig:qmc_gates}). In this case, the performance of QMC would be much better, making gate times as slow as $1000$ns competitive against most HPC systems. For gate times faster that $10$ns, even the supercomputer would be slower than the quantum computer for calculations taking longer than two-and-a-half hours---in the case of $10$ns gate times---and longer than a minute-and-a-half---in case of $1$ns gate times. Whether it is possible to implement QMC with such low-depth quantum circuits as in Figure \ref{fig:qmc_gates} remains to be seen, but the possibility is not unrealistic. Ideas for how to achieve this will be discussed in the next section.

\section{Discussion: Challenges and Future Directions of Research}\label{sec:extensions}

The previous section demonstrates that QMC can be advantageous over sizable HPC clusters. However, for QMC to outperform supercomputers in terms of computational speed on realistic problems, much progress will be needed on both the hardware and the algorithmic fronts. In terms of quantum hardware, fast gate times will be essential to realize the quantum speedup in practice (see Figure \ref{fig:qmc_time}). Moreover, the coherence times (see section \ref{Sec:q_hardware}) will need to be longer than the duration of the computation, lest the result be reduced to noise. Also, the number of qubits will need to rise significantly, keeping their error rates below the fault-tolerance threshold (see section \ref{sec:fault_tol}). This way, through error-correction, the errors from quantum noise will not interfere with the computation.

However, significant progress will also need to occur within the QMC algorithm itself, reducing the depth (the number of quantum gates that need to be executed sequentially) of the quantum circuit. One problem is that arithmetic (multiplication and especially addition) is costly on a quantum computer. For this, efficient ways of performing quantum arithmetic are being developed (see, for example, \citealp{li2020efficient,ruiz2017quantum}). Another idea for improving QMC lies in modifying the amplitude estimation part of the algorithm (the part that gives rise to the quadratic speedup over classical MC). This has been investigated in \cite{suzuki2020amplitude,burchard2019lower,grinko2019iterative,giurgica2020low}.

Another challenge is how to encode the classical problem efficiently onto the quantum computer. In this paper, we use the approximate scheme of \cite{egger2020credit} and \cite{ Stamatopoulos_2020}, but this reduces the scaling exponent (for the power law dependence of precision vs oracle calls) from faster ${-1}$ to slower ${-\frac{2}{3}}$ . One alternative is in \cite{herbert2021quantum} which retains the $-1$ scaling exponent by performing the Fourier series decomposition of the sum that approximates the MC integral. An altogether different approach would be to try encoding the classical problem with a variational circuit. We have used a variational circuit to encode the probability distribution (see unitary $\mathcal{A}$). Extending this to encode the entire problem will likely require overcoming more technical challenges.\footnote{For example the barren plateau problem (\citealp{uvarov2021barren}).} Nevertheless, given a variational circuit's very low depth, this is a worthwhile endeavor to consider.

Finally, it should be noted that lessons from classical computational methods can be used to improve the QMC algorithm, as well. For instance, to reduce discretization error, rather than considering the entire probability distribution, it is possible to sample just from the ergodic set (\citealp{oxtoby1952ergodic}). Implementing ideas from importance sampling (\citealp{tokdar2010importance}) could also be beneficial.

\section{Conclusion}\label{conclusion}

This paper is the first to apply the quantum Monte Carlo algorithm to problems in economics and among the first to apply quantum computation more generally to this field. We build our paper in such a way that an economist without any knowledge of quantum computation can gradually progress to fully implementing the QMC algorithm. We begin with a simple Gaussian sampling needed to estimate the mean of a random variable to illustrate quantum decomposition. Next, we focus on more sophisticated applications: (a) stress testing of banks and (b) solving DSGE models with deep learning. We derive in detail how these problems can be converted into quantum circuits---both theoretically and in terms of code. The quantum simulations that we perform for these problems verify that quantum solutions converge to theoretical predictions. In the absence of hardware noise, high accuracy is achieved even for a small number of qubits.

The other important contribution of this paper is a fair comparison between QMC and classical MC. Whereas QMC speedup is usually represented in terms of the number of oracle calls (number of times the unitary $\mathcal{Q}$ is applied), this paper shows how to obtain a direct time-vs-time comparison. For the simple DSGE deep learning problem, we decompose the quantum circuit into a set of standard elementary gates. This allows us to compute the depth of the circuit, which we convert into computational time of the QMC algorithm for various quantum gate times. The resulting graph informs when quantum advantage might be expected for this economics problem.

We also consider a more realistic consumption-saving macroeconomic model with an occasionally binding borrowing constraint and exogenous stochastic shocks. We show the difficulty in solving this model accurately with deep learning. We demonstrate that QMC retains the quadratic speedup for finding the precise mean-squared error of the computed Bellman equation. This, in turn, suggests that QMC could one day increase the convergence rate of the deep learning method and, in turn, allow for more accurate solutions of economics models.

With regard to the QMC algorithm itself, this paper makes two contributions. We show how the unitary $\mathcal{A}$ encoding the probability distribution of the random variable can be approximated with a quantum variational circuit. Also, we show how the controlled $\mathcal{Q}$ (quantum oracle) unitary can be achieved without the controlled $\mathcal{F}$ unitary (which encodes the classical problem onto the quantum circuit.

Although further improvements in quantum hardware and the QMC algorithm itself are necessary before QMC can achieve an advantage over classical MC in practice, this paper shows that QMC scales better than classical MC for problems in economics. Therefore, especially in cases where renting a supercomputer is not realistic, quantum Monte Carlo stands to deliver speed gains in solving economic problems.

\section*{ACKNOWLEDGEMENTS}

We thank Kenneth Judd, Jonathan Chiu, Matias Vieyra, Sepehr Taghavi, and Juan Miguel Arrazola,  for fruitful discussions.

The views expressed in this paper are solely those of the authors and may differ from official Bank of Canada views. No responsibility for them should be attributed to the Bank.

\bibliography{references.bib}

\newpage 
\appendix

\section{Optimizing a variational circuit to prepare the probability distribution}\label{app:a_train}

Section \ref{Sec:A_decomp} discusses how a variational quantum circuit can be optimized to
output a target probability distribution when sampled in the computational basis. The code
below shows how this can be achieved using the probability distribution and $m=5$-qubit
setting detailed in \ref{Sec:Example}.

\custompythonfromfile{./code_fragments/a_train.py}

The variational circuit is composed of $5$ layers. Each layer applies a single-qubit rotation
gate to each qubit, with each gate having $3$ parameters allowing for a general rotation in
the Bloch sphere. The rotation gates are followed by an entangling block of CNOT gates.

\begin{figure}[h]
\includegraphics[width=0.65\textwidth]{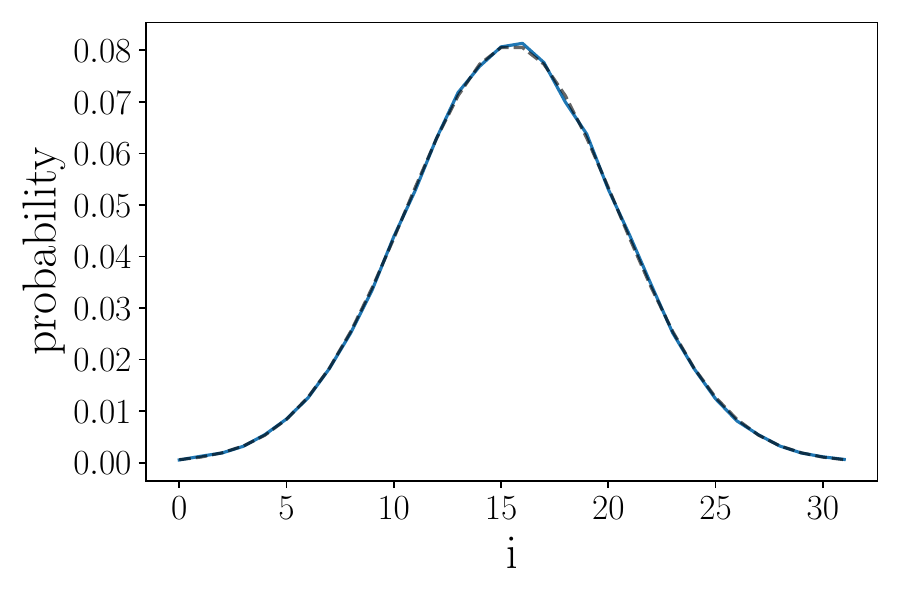}
\caption{Optimizing a variational quantum circuit so that its output probability distribution
in the computational basis is the discretized normal distribution. The solid blue line shows
the output distribution from the optimized circuit, while the dashed grey line shows the
target distribution.}
\label{fig:a_train}
\end{figure}

When this code is used to reproduce the discretized normal distribution discussed
in section \ref{Sec:Example}, the resulting probability distribution $p(i, \bm{\theta}_{\rm opt})$ is shown in Figure~\ref{fig:a_train}.
.

\section{Discretizing the simple example}\label{app:simple_example}

The following code shows how the simple example in section \ref{Sec:Example} can be
discretized using Python code.

\custompythonfromfile{./code_fragments/qmc.py}

\section{Linear approximation of the $\mathcal{R}$ unitary which encodes the random variable}\label{app:linear_R}

\begin{figure}[t]
\mbox{
\Qcircuit @C=0.6em @R=0.5em {
& \qw & \qw & \qw & \qw & \qw & \qw & \qw & \ctrl{5} & \qw \\
& \vdots  & & & & & & & \\
& & & & & & & \\
& \qw & \qw & \qw  & \ctrl{2} & \qw & \qw & \qw & \qw & \qw \\
& \qw & \qw & \ctrl{1} & \qw & \qw & \qw & \qw & \qw & \qw \\ 
& \gate{R_{Y}(\pi / 4)} & \gate{R_{Y}(b)} & \gate{R_{Y}(a)} & \gate{R_{Y}(2 a)} & \qw & \cdots & & \gate{R_{Y}(2^{m - 1} a)} & \qw
}
}
\caption{Encoding a linear function $f(x)$ onto an ancilla qubit from an $m$-qubit
register can be achieved by discretizing and rescaling to
$f(i) = a i + b$ and applying the above
circuit~\cite{egger2020credit}. Here,
$R_{Y}(\phi) = e^{- i \phi \sigma_{Y}}$ with $\sigma_{Y}$ the Pauli-Y operator.}
\label{fig:CRYcascade_linear}
\end{figure}
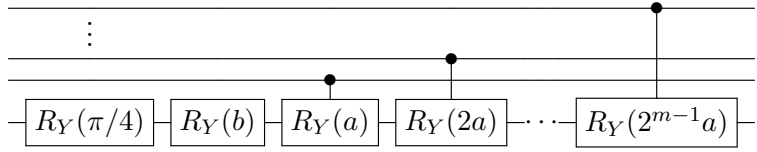

To implement the linear approximation of $\mathcal{R}$ in \cite{egger2020credit, Stamatopoulos_2020} as a one-dimensional linear function $f(x)$, the function should first
be discretized and rescaled to the range $[-c_s, c_s]$ for some small
constant $c_s>0$. The result can be written as
\begin{equation}
    f(i) = a i + b,
\end{equation}
with two constants $a$ and $b$ that depend on $c_s$. The cascade of controlled-Y
rotations shown in Figure~\ref{fig:CRYcascade_linear} can then be
shown to approximate $\mathcal{R}$, that is, so that the probability
of measuring the ancilla qubit in the $\ket{1}$ state can be
rescaled according to $c_s$ to provide an estimate of
${\rm E}(f(x))$. This holds provided that $c_s$ is sufficiently
small, such that $\sin^{2}(f(i) + \pi / 4) \approx f(i) + 1/2$. 
Indeed, $c_s$ must be set so that the dominant error is from the Monte
Carlo estimation, and it can be shown that the optimal $c_s$ scales
with the number $N$ of applications of $\mathcal{Q}$
as $N^{- 1 / 3}$ (\citealp{woerner2019quantum}). However, this approximation results in a less preferential scaling with
$N = \mathcal{O}(\varepsilon^{-3/2})$.

\section{Controlled $\mathcal{Q}$ without controlled $\mathcal{F}$}\label{app:Q_decomp}

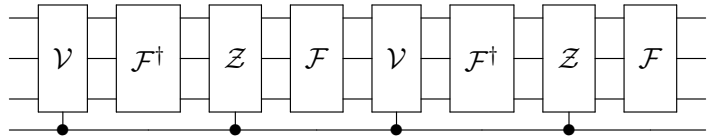
\begin{figure}[h]
\mbox{
\Qcircuit @C=1em @R=0.5em {
& \multigate{2}{\mathcal{V}} & \multigate{2}{\mathcal{F}^{\dagger}} & \multigate{2}{\mathcal{Z}} & \multigate{2}{\mathcal{F}} & \multigate{2}{\mathcal{V}} & \multigate{2}{\mathcal{F}^{\dagger}} & \multigate{2}{\mathcal{Z}} & \multigate{2}{\mathcal{F}} & \qw \\
& \ghost{\mathcal{V}} & \ghost{\mathcal{F}^{\dagger}} & \ghost{\mathcal{Z}} & \ghost{\mathcal{F}} & \ghost{\mathcal{V}} & \ghost{\mathcal{F}^{\dagger}} & \ghost{\mathcal{Z}} & \ghost{\mathcal{F}} & \qw \\
& \ghost{\mathcal{V}} & \ghost{\mathcal{F}^{\dagger}} & \ghost{\mathcal{Z}} & \ghost{\mathcal{F}} & \ghost{\mathcal{V}} & \ghost{\mathcal{F}^{\dagger}} & \ghost{\mathcal{Z}} & \ghost{\mathcal{F}} & \qw \\
& \ctrl{-1} & \qw & \ctrl{-1} & \qw & \ctrl{-1} & \qw & \ctrl{-1} & \qw & \qw \\
}
}
\caption{A controlled version of $\mathcal{Q}$ can be achieved without using a controlled-$\mathcal{F}$.}
\label{fig:Q_cntrl}
\end{figure}

Figure~\ref{fig:Q_cntrl} shows the quantum circuit for achieving $\mathcal{Q}$ with control from the phase estimation qubit without the need for a controlled version of $\mathcal{F}$.

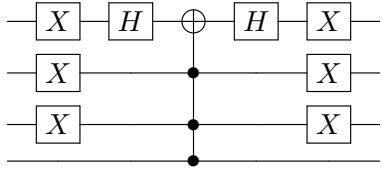
\begin{figure}[h]
\mbox{
\Qcircuit @C=1em @R=0.5em {
& \gate{X} & \gate{H} & \targ & \gate{H} & \gate{X} & \qw \\
& \gate{X} & \qw & \ctrl{-1} & \qw & \gate{X} & \qw \\
& \gate{X} & \qw & \ctrl{-1} & \qw & \gate{X} & \qw \\
& \qw & \qw & \ctrl{-1} & \qw & \qw & \qw \\
}
}
\caption{Applying a controlled version of $\mathcal{V}$ can be achieved using Pauli-X rotations, Hadamard gates, and a multi-controlled-X gate. Note that the target for the
multi-controlled-X gate can be any of the wires in the $m$-qubit register, provided the
Hadamard gates are also applied to that wire.}
\label{fig:Z_ctrl}
\end{figure}

Let us now consider the unitary $\mathcal{V}$. It can be seen that
\begin{align}
\mathcal{V} =& \,\,\mathbbm{1}_{m+1} - 2 \mathbbm{1}_{m} \otimes |1\rangle\langle 1| \notag \\
=& \,\,\mathbbm{1}_{m} \otimes \sigma_{\rm z},
\end{align}
with $\sigma_{\rm z}$ the Pauli-Z matrix. Hence, controlled-$\mathcal{V}$ can be implemented
simply as a CZ gate. Furthermore, the controlled-$\mathcal{Z}$ gate can be implemented as shown
in Figure~\ref{fig:Z_ctrl}. This requires the ability to apply a multi-controlled-X gate,
for which an efficient decomposition exists that scales linearly with
$m$ (\citealp{barenco1995elementary}).

\section{Details of the QMC implementation for the neoclassical model}\label{app:QCMNeoclassical}

We discretize our target expression (\ref{optimalSol}) into $M=32$ points using $m=5$ qubits and rescale
the function to $f(i) = a i + b \in [-c_s, c_s]$ according to a
constant $c_s>0$, such that $a = 2c_s / 31$ and $b = -c_s$. When using
$n$ qubits and $N=2^n$ applications of $\mathcal{Q}$ in the QMC
algorithm, we set $c_s = \sqrt[3]{3\pi/N}$ as suggested in \cite{woerner2019quantum}.

Using this,
$\mathcal{R}$ can be applied according to the cascade of controlled-Y
rotations in Figure~\ref{fig:CRYcascade_linear}. For the
probability-encoding unitary $\mathcal{A}$, we set $x_{i}$ and
$p(i)$ using equation \eqref{eq:disc_ex} with $x_{\max} = 1.06$ and $x_{\min} = 0.94$ (which corresponds to three standard deviations away from the mean), and
carry out the optimization procedure in section \ref{Sec:A_decomp}.
With $10$ trainable layers, $150$ parameters are varied to minimize
the cost function $C(\bm{\theta})$ in equation \eqref{Eq:Cost}. The training was carried out with the learning rate equal to $0.01$ for the first $1000$ epochs, $0.001$ for the next $1000$ epochs, and $0.0001$ for the final $1000$ epochs (see \ref{app:a_train} for details). The final cost reached this way is $1.54 \times 10^{-4}$.
The code block below shows how the resulting $\mathcal{F}$ can
be constructed in PennyLane.
\begin{custompython}
m = 5
c_s = np.power(3*np.pi/N, 1 / 3)
a = 2 * c_s / 31
b = -c_s
ancilla_qubit = m
a_qubits = range(m)

def F():
    A(theta)  # Apply A trained previously

    # Apply R
    qml.RY(np.pi / 2, wires=ancilla_qubit)
    qml.RY(2 * b, wires=ancilla_qubit)
    for qubit in a_qubits:
        phi = 2 ** (m - qubit) * a
        qml.CRY(phi,wires=[qubit,ancilla_qubit])
\end{custompython}
%        qubit_pair = [qubit, ancilla_qubit]

With the components of $\mathcal{F}$ fixed, we can then proceed
to apply the QMC algorithm. This can be achieved
as follows (where “dev" can refer to a quantum device or a classical simulator):
\begin{custompython}
#n is number of estimation qubits chosen
est_qubits = range(m + 1, n + m + 1)
@qml.qnode(dev)
def qmc():
    qml.quantum_monte_carlo(
        F, qubits, ancilla_qubit, est_qubits)
    return qml.probs(est_qubits)
\end{custompython}

The quantum resource requirements of this circuit can be obtained as a Python dictionary using the following code:
\begin{custompython}
expanded_tape = qmc.qtape.expand(depth=10)
specs = expanded_tape.specs
\end{custompython}

\end{document}